\documentclass[preprint,showpacs,preprintnumbers,amsmath,amssymb]{revtex4}
\usepackage{amsmath,amssymb,graphics,epsfig,subfigure,color}
\usepackage[subfigure]{graphfig}

\begin{document}
\newcommand {\nn}    {\nonumber}

\title{Resonance mass spectra of gravity and fermion on Bloch branes}

\author{Qun-Ying Xie$^{1,2}$\footnote{xieqy@lzu.edu.cn}\footnote{corresponding author},
        Jie Yang$^{2}$,
        Li Zhao$^{2}$}
\affiliation{
  $^{1}$School of Information Science and Engineering, Lanzhou University, Lanzhou 730000, People's Republic of China\\
  $^{2}$Institute of Theoretical Physics, Lanzhou University, Lanzhou 730000, People's Republic of China
          }

\begin{abstract}
In this paper, by presenting the
potentials of Kaluza--Klein (KK) modes in the corresponding
Schr\"{o}dinger equations, we investigate the localization and resonances of
gravity and fermion on the symmetric and asymmetric Bloch {branes}.
We find that the localization properties of zero modes for gravity and fermion in the symmetric
brane case are the same, whereas, for the asymmetric brane case, the fermion zero mode
is localized on one of the sub-branes, while the gravity zero mode is localized
on another sub-brane. The
spectra of the gravity and the left- or right-handed fermion are composed of
a bound zero mode and a series of gapless continuous massive KK
modes. Among the continuous massive KK
modes, we obtain some discrete gravity and fermion resonant (quasilocalized) KK states on the brane, which
have a finite probability of escaping into the bulk. The KK
states with lower resonant masses have a longer lifetime on the brane.
And the number of the resonant KK states increases linearly with
the width of the brane and the scalar-fermion coupling constant, {but it  decreases with the asymmetric factor $\beta$}.
The structure of the resonance spectrum is investigated in detail.
\end{abstract}




\maketitle

\section{Introduction}

About ninety years ago, the idea of the possible existence of extra spatial dimensions was presented by Kaluza and Klein (KK) \cite{Kaluza1921,Klein1926}, who tried to
unify four-dimensional (4D) Einstein gravity and electromagnetism by proposing a theory with a compact fifth dimension. However, the KK theory is not
a viable model to describe nature because it has many problems. Later, in 1980s, our 4D universe was considered as a domain wall (topological defect) embedded in a higher-dimensional space-time \cite{Rubakov1983,Akama1983,Antoniadis1990}. {After large \cite{ADD} and warped \cite{rsI} extra dimensions were respectively presented as a solution to the hierarchy problem, the idea of extra dimensions became popular.} In this scenario, our 4D universe is an infinitely thin brane embedded in a higher dimensional space-time, and gravity is free to propagate in all dimensions, while all the Standard Model fields are localized on a 3-brane.
{In the Randall-Sundrum-1 (RS1) model \cite{rsI}, there are two 3-branes located at the boundaries of a compact extra dimension with the topology $S^1/Z_2$, one with negative tension (the visible brane or TeV brane we lived on) and another with positive tension (the hidden brane or Planck brane). In this model, the gauge hierarchy problem is solved by an exponential warp factor $e^{-ky}$, but there is a modulus problem and we will get a ``wrong-signed" Friedmann-like equation on our negative tension brane. Hence, in order to stabilize the modulus, one needs to introduce a scalar field in the bulk \cite{GW}. In the RS2 model \cite{rsII}, the extra dimension is infinite and there is only one brane with positive tension, on which gravity can be localized. The RS2 model does not suffer from the modulus problem and can reduce a right-signed Friedmann-like equation on the brane, but the hierarchy problem is left.}
Generalizations and extensions of the RS brane model can be found
for examples in Refs.
\cite{ExtensionRS,littleStringBrane,1104.3188,scalar-tensor branes}.

{In this paper, we are interested in the generalization of the RS2 model, which is done by introducing background scalar fields in the bulk. In this generalization, bulk scalars play the role of generating the brane as a domain wall (thick brane).
And a virtue is that branes can be realized naturally and have inner structure.} Because of this, an increasing interest has
been focused on investigation of thick branes generated by bulk scalar fields in higher dimensional space-time
\cite{dewolfe,GremmPLB2000,Csaki,ThickBrane4,Liu0907.1952,George2010a,George2010b,Zhao2010,1004.2181,Zhong2011,ThickBraneBazeia}.
For a review on thick brane solutions see Ref. \cite{ThickBraneReview}.


Localization and spectra of various bulk fields including gravity on a brane is an important and interesting problem
\cite{ThickBrane4,RandjbarPLB2000,Volkas2007,Liu0708,20082009,Fu2011,Liu0803,LiLiu2011,Gogberashvili2012,Liu2012a,Christiansen2012}.
Because the effective physics in the low energy scale is four-dimensional, the lowest modes of various bulk fields should be localized on the brane in order not to contradict the current experiments.
In order to localize fermions on branes, one needs to introduce localization
mechanisms \cite{RandjbarPLB2000}. Usually the Yukawa coupling of fermions with background scalar fields \cite{Volkas2007} is introduced. Under this mechanism, there may exist a
single bound massless fermion KK mode (i.e., the fermion zero mode)
\cite{Liu0708,20082009,Fu2011}, or a bound massless KK mode and finite discrete bound massive KK modes
(mass gaps) \cite{ThickBrane4,Liu0803}.
Furthermore, some other
backgrounds, for example, supergravity \cite{Mario} and gauge field
\cite{Parameswaran0608074} could be also considered.

Fermions and gravity on symmetric double branes have been investigated
in Refs. \cite{Guerrero2006,MelfoPRD2006}. There are two sub-branes located at the edges of the double brane. It was shown that 
the fermion zero mode on the double brane is not peaked at the center of
the brane, but instead is a constant between the two sub-branes.

In Ref.~\cite{BazeiaJHEP2004}, a kind of double brane, the so-called Bloch brane, was presented by Bazeia and Gomes. In this model, the system is described by two real scalar fields coupled with gravity in warped space-time. It was found that the parameter which controls the way the two scalar fields interact results in the appearance of a thick double brane with internal structure. In this paper, we would like to investigate the localization of fermion and gravity on the Bloch brane. Especially, we will find that there are many resonant KK modes for gravity and fermion on the double brane, which in fact are quasi localized KK modes. The resonant KK modes with lower resonant mass have a longer lifetime. We also construct an asymmetric Bloch brane and investigate localization of zero modes of gravity and fermion on it. It will be shown that the fermion zero mode is localized on one of the sub-branes, while the gravity zero mode is localized on another sub-brane. As far as we know, the resonant spectra of gravity and fermion on asymmetric branes still has not been investigated. Therefore, the resonant structure for gravity and fermion on the asymmetric Bloch brane will also be investigated in this paper.

The paper is organized
as follows. In Sec.~\ref{SecModel}, we first give a brief review of
the symmetric Bloch brane in five-dimensional space-time, and construct asymmetric Bloch brane solutions. Then, in Sec.~\ref{SecGravityLocalize}, we study the zero mode and resonance mass spectrum
of gravity on the symmetric and asymmetric thick brane by presenting the
potential of the corresponding Schr\"{o}dinger problem for the linear tensor perturbation KK modes of the metric. 
In Sec.~\ref{SecFermionLocalize}, we investigate the localization and resonance mass spectrum
of spin-half fermions on the symmetric and asymmetric thick brane with the typical
scalar-fermion interaction.
We simply
compare the localization of the zero modes of gravity
and fermion on the symmetric and asymmetric branes in Sec.~\ref{secStrongWeakBranes}. Finally, a
brief discussion and conclusion are presented in Sec.
\ref{secConclusion}.

\section{Review of the model}
\label{SecModel}

Let us consider thick branes arising from two interacting real
scalar fields $\phi$ and $\chi$ with a scalar potential
$V(\phi,\chi)$. The action for such a system is given by
\begin{equation}
S = \int d^5 x  \sqrt{-g}\left [ \frac{1}{4} R
 -\frac{1}{2} \left(\partial^M \phi \partial_M \phi
                         +\partial^M \chi \partial_M \chi \right)
 - V \right], \label{action}
\end{equation}
where the five-dimensional gravitational
constant is chosen as $G^{(5)}=1/(4\pi)$, and $V=V(\phi,\chi)$ is the scalar potential of $\phi$ and $\chi$. The line element for a
five-dimensional space-time describing a {{Minkowski}} brane is assumed
as
\begin{eqnarray}
 ds^2&=&\text{e}^{2A(y)}\eta_{\mu\nu}dx^\mu dx^\nu+ dy^2, \label{linee}
\end{eqnarray}
where $\text{e}^{2A(y)}$ is the warp factor. The scalar fields are
considered to be functions of $y$ only for the static {{Minkowski}} brane scenario, i.e., $\phi=\phi(y)$ and
$\chi=\chi(y)$. In the model, the potential could provide a thick
brane realization, and the brane configuration with warped geometry
would have a nontrivial interior structure. The field equations
generated from the action (\ref{action}) under these assumptions
reduce to the following set of second-order nonlinear coupled
differential equations
\begin{subequations}\label{2orderEqs}
\begin{eqnarray}
 \frac{d^2\phi}{dy^2} + 4 \frac{dA}{dy} \frac{d\phi}{dy}
      & = & \frac{\partial V(\phi,\chi)}{\partial\phi}, \\
 \frac{d^2\chi}{dy^2} + 4 \frac{dA}{dy} \frac{d\chi}{dy}
      & = & \frac{\partial V(\phi,\chi)}{\partial\chi},\\
 \left(\frac{dA}{dy}\right)^2 +\frac{1}{4} \frac{d^2A}{dy^2}& = &
     -\frac{1}{3} V(\phi,\chi),\\
  \left(\frac{d\phi}{dy}\right)^2
     +\left(\frac{d\chi}{dy}\right)^2
     & = &- \frac{3}{2}\frac{d^2A}{dy^2}.
\end{eqnarray}\end{subequations}

It is useful to reduce the above Einstein-scalar equations to the
first-order equations by introducing an auxiliary superpotential
$W=W(\phi,\chi)$ \cite{dewolfe,GremmPLB2000,BazeiaJHEP2004,SuperPotential,DutraPLB2005}.
Provided the potential
\begin{eqnarray}
 V&=&\frac{1}{2}
  \left[ \left( \frac{\partial W}{\partial \phi}\right)^2
         +\left( \frac{\partial W}{\partial \chi}\right)^2 \right]
  -\frac{4}{3} W^2,
 \label{Vphichi_W)}
\end{eqnarray}
the above second-order equations (\ref{2orderEqs}) become
\begin{subequations}\label{1orderEqs}\begin{eqnarray}
 \frac{d\phi}{dy} & = & \frac{\partial W}{\partial\phi}, \\
 \frac{d\chi}{dy} & = & \frac{\partial W}{\partial\chi},\\
 \frac{dA}{dy}& = & -\frac{2}{3} W.
\end{eqnarray}\end{subequations}
These equations would be helpful to give analytical brane solutions.

\subsection{Symmetric Bloch brane}

For the superpotential $W$ given by
\begin{eqnarray}
 W = \phi\left[\left(1-\frac{1}{3}\phi^2\right)
                -b\chi^2\right],\label{BlochW1}
\end{eqnarray}
the symmetric Bloch brane solution was found in Ref.
\cite{BazeiaJHEP2004}:
\begin{subequations}\label{BlochBrane1}\begin{eqnarray}
 \phi(y) &=& \tanh (2by),\label{BlochBrane1Phiy}\\
 \chi(y) &=& \sqrt{\left(\frac{1}{b}-2\right)}
             ~\text{sech} (2by),\label{BlochBrane1Chiy}\\
 A(y) &=& \frac{1}{9b}\bigg[ (1-3b)\tanh^2 (2by)
       -2\ln \cosh(2by) \bigg].\label{BlochBrane1Ay}
\end{eqnarray}
\end{subequations}
where the parameter $b$ satisfies $0<b<1/2$. Notice that the limit
$b\to1/2$ changes the two-field solution to the one-field solution.
The two-field solution represents a Bloch wall, while the one-field
solution is an Ising wall.

\begin{figure}[htb]
\begin{center}
\includegraphics[width=7cm]{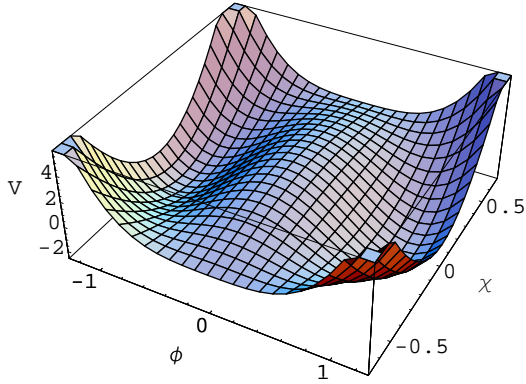}
\end{center}
\vskip -4mm \caption{The shape of the scalar potential
$V(\phi,\chi)$ for symmetric brane scenario. The parameters are set
to $a=b=2$ and $v=1$.}
 \label{fig_Vphichi}
\end{figure}

The generalized superpotential
\begin{eqnarray}
 W=\phi\left[ a\left(v^2-\frac{1}{3}\phi^2\right)
              -b\chi^2\right] \label{BlochW2}
\end{eqnarray}
was considered in Ref. \cite{DegenerateBlochbranes}. Figure. \ref{fig_Vphichi}
shows the shape of the scalar potential $V(\phi,\chi)$ with the parameters $a=b=2,~v=1$. It can be seen that there are two vacua in this scalar potential, which are located at $\chi=0,~\phi=\pm v$. The following general Bloch
brane solution was found for the cases $a>2b>0$ and $a<2b<0$ in Ref.
\cite{DegenerateBlochbranes}:
\begin{subequations}\label{BlochBrane2}
\begin{eqnarray}
 \phi(y) &=& v \tanh (2bvy),\label{BlochBrane2Phiy}\\
 \chi(y) &=& v \sqrt{\frac{a-2b}{b}}~\text{sech} (2bvy),\label{BlochBrane2Chiy}\\
 A(y) &=& \frac{v^2}{9b}
    \bigg[ (a-3b)\tanh^2 (2bvy)
           -2a \ln \cosh(2bvy) \bigg].\label{BlochBrane2Ay}
\end{eqnarray}
\end{subequations}
Note that by setting $a=v=1$, the special superpotential
(\ref{BlochW1}) and the Bloch brane solution (\ref{BlochBrane1}) are
recovered.

It is very interesting that there exist degenerate Bloch brane
solutions for $a=b$ and $a=4b$
\cite{DutraPLB2005,DegenerateBlochbranes}. For the case $a=b$, the
solution is \cite{DegenerateBlochbranes}
\begin{subequations}\label{DBbrane1}
\begin{eqnarray}
 {\phi}(y)
   &=&  v\frac{\sqrt{c_{0}^{2}-4}\sinh (2bvy)}
              {\sqrt{c_{0}^{2}-4}\cosh (2bvy)-c_{0}},
              \label{DBbrane1phi} \\
 {\chi}(y)
   &=& \frac{2v}{\sqrt{c_{0}^{2}-4}\cosh (2bvy)-c_{0}},
              \label{DBbrane1chi} \\
 e^{2A(y)}
  &=&\bigg( \frac{\sqrt{c_{0}^{2}-4}-c_{0}}
         {\sqrt{c_{0}^{2}-4}\cosh (2bvy)-c_{0}}\bigg)
          ^{4v^{2}/9} \nonumber \\
  &\times&\exp \bigg[ -\frac{4v^2\big( c_{0}^{2}-4-c_{0}
     \sqrt{c_{0}^{2}-4}\big) }
     {9\big( \sqrt{c_{0}^{2}-4}
     -c_{0}\big) ^{2}}\bigg]\nonumber \\
  &\times& \exp \bigg[ \frac{4v^2\big( c_{0}^{2}-4-c_{0}
     \sqrt{c_{0}^{2}-4}\cosh(2bvy)\big) }
     {9\big( \sqrt{c_{0}^{2}-4}\cosh(2bvy)
     -c_{0}\big) ^{2}}\bigg],\label{DBbrane1E2A}
\end{eqnarray}\end{subequations}
where $c_{0}<-2$. For $a=4b$, one can get
\cite{DegenerateBlochbranes}
\begin{subequations}\label{DBbrane2}
\begin{eqnarray}
 {\phi}(y)
   &=&  \frac{v\sqrt{1-16c_0}~\text{sinh}(4 b v y)}
              {1+\sqrt{1-16c_0}~\cosh(4 b v y)},
             \label{DBbrane2phi} \\
 {\chi}(y)
   &=& \frac{2v}{\sqrt{1+\sqrt{1-16c_0}~\cosh(4 b v y)}},
              \label{DBbrane2chi}\\
 e^{2A(y)}
  &=& \left(\frac{1+\sqrt{1-16 c_0}}
                 {1+\sqrt{1-16 c_0}\cosh[4bvy]}
      \right)^{\frac{8v^2}{9}}\nonumber\\
  &\times&  \exp\left[\frac{{4v^2}\big(1+8 c_0+\sqrt{1-16c_0}\big)}
                           {9\big(1+\sqrt{1-16 c_0}\big)^2}\right]\nonumber\\
  &\times&  \exp\left[-\frac{{4v^2}\big(1+8 c_0+\sqrt{1-16c_0}
        \cosh(4 b v y)\big)}{9\big(1+\sqrt{1-16 c_0}
        \cosh(4 b v y)\big)^2}\right],\label{DBbrane2E2A}
\end{eqnarray}\end{subequations}
where $c_0<1/16$. The above two brane solutions (\ref{DBbrane1}) and (\ref{DBbrane2}) have similar properties, so we only focus on the solution (\ref{DBbrane1}) in this paper.

%
%
%
%
%
%

\subsection{Asymmetric Bloch brane}

Next, we construct asymmetric Bloch brane solutions by shifting the
superpotential (\ref{BlochW2}) a positive constant $W \rightarrow
W+3\beta/2$. Note that the shift does not change the solutions
for the scalars $\phi$ and $\chi$. The extrema of the new scalar
potential are also at $\chi=0,~\phi=\pm v$. Now, the form of the
warp factor exponent is changed to $A(y)-{\beta}y$ with $A(y)$ given
by (\ref{BlochBrane1Ay}),~(\ref{BlochBrane2Ay}),
~(\ref{DBbrane1E2A}),~(\ref{DBbrane2E2A}) for the four symmetric
Bloch branes above, respectively. In order to have a finite value
for $A(y)-{\beta}y$ at $y\rightarrow\pm\infty$, we need some limit
on the parameter $\beta$, which turns out to be
\begin{eqnarray}
  | \beta | \leq\frac{4}{9}av^3
\end{eqnarray}
for all the solutions. The space-time far away from the brane is
AdS$_5$ with different cosmological constants:
\begin{eqnarray}
 \Lambda_{\pm}=-\frac{1}{27}(9\beta\pm4bv^3)^2.
\end{eqnarray}
This is similar to the situation in Refs. \cite{Guerrero2006,MelfoPRD2006}.
While the cosmological constant at the origin of the extra dimension
has a slightly different expression:
\begin{eqnarray}
 \Lambda_{0}=\left\{\begin{array}{ll}
     \frac{1}{2}(a-3b)^2 v^4 -3\beta^2
           & ~~\text{for~solution}~(\ref{BlochBrane2})  \\
     \frac{1}{2}b^2v^4 \big(1-\frac{4}{(c_0-\sqrt{c_0^2-4})^2}\big)^2 -3\beta^2
           & ~~\text{for~solution}~(\ref{DBbrane1}) \\
     \frac{1}{2}b^2v^4 \big(1-\frac{4}{1+\sqrt{1-16c_0}}\big)^2 -3\beta^2
           & ~~\text{for~solution}~(\ref{DBbrane2})
             \end{array}\right.  ,
\end{eqnarray}
which is positive for a symmetric brane scenario and can be positive,
zero, or negative for an asymmetric scenario.

\begin{figure}[htb]
\begin{center}
\includegraphics[width=7cm,height=4.5cm]{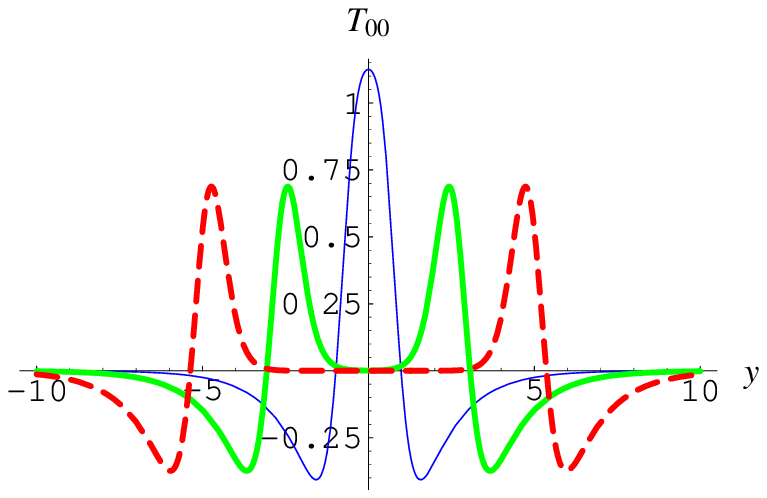}
\includegraphics[width=7cm,height=4.5cm]{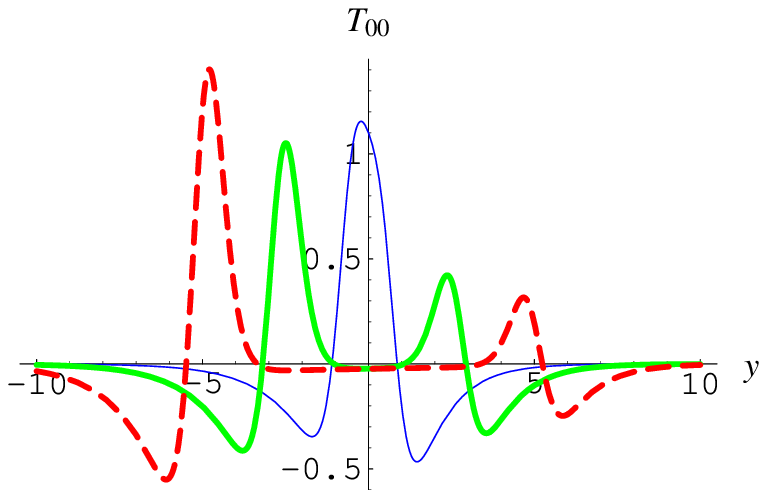}
\end{center}
\vskip -4mm
 \caption{The shape of the energy density ($T_{00}$) for the symmetric
 ($\beta=0$, left) and asymmetric ($\beta=1/16$, right)
 degenerate Bloch branes with $a=b$.
 The parameters are set to $b=1,~v=1$,
 $c_0=-2- 10^{-8}~(\delta_0=9.9)$ for the red dashed lines,
 $c_0=-2- 10^{-4}~(\delta_0=5.3)$ for the green thick lines, and
 $c_0=-2.5$ for the blue thin lines.}
 \label{fig_EnergyDensity_DB1}
\end{figure}

\begin{figure}[htb]
\begin{center}
\includegraphics[width=7cm,height=4.5cm]{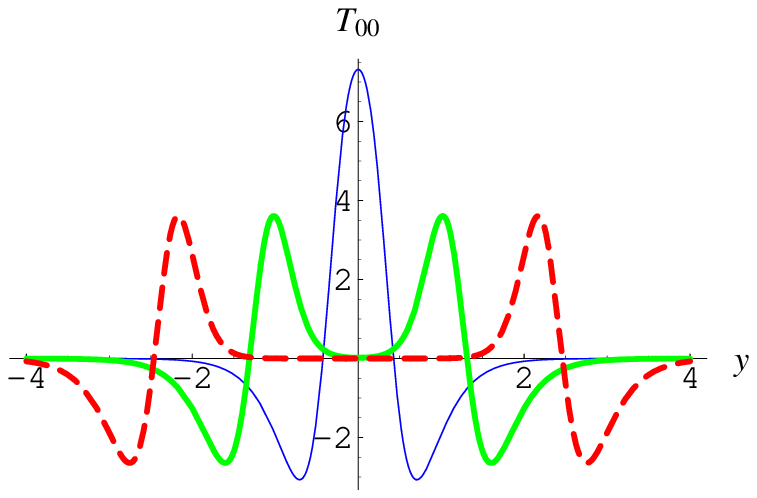}
\includegraphics[width=7cm,height=4.5cm]{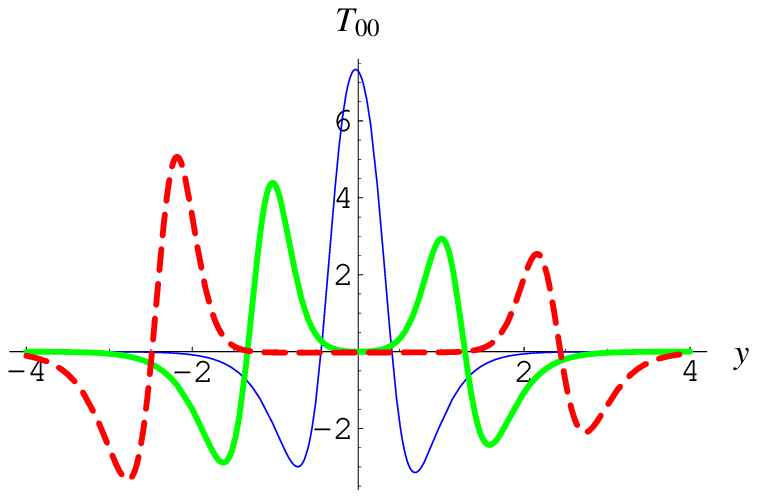}
\end{center}\vskip -4mm
 \caption{The shape of the energy density for the symmetric
 ($\beta=0$, left) and asymmetric ($\beta=1/16$, right)
 degenerate Bloch branes with $a=4b$.
 The parameters are set to $b=1,~v=1$,
 $c_0=\frac{1}{16+10^{-6}}(\delta_0=5.0)$ for the red dashed lines,
 $c_0=\frac{1}{16+10^{-2}}(\delta_0=2.7)$ for the green thick lines, and
 $c_0=0.01$ for the blue thin lines.}
 \label{fig_EnergyDensity_DB2}
\end{figure}

The shape of the energy density $\rho$ for the symmetric
and asymmetric degenerate Bloch branes is shown in Figs.
\ref{fig_EnergyDensity_DB1} and \ref{fig_EnergyDensity_DB2}.
The thickness of the degenerate branes could be estimated as
\begin{eqnarray}
 \delta=\left\{\begin{array}{ll}
     \frac{1}{b v} \ln \frac{-2c_0}{\sqrt{c_0^2-4}}
           & ~~\text{for~solution}~(\ref{DBbrane1})
             ~\text{with}~c_0\rightarrow-2 \\
     \frac{1}{2b v} \ln \frac{6}{\sqrt{1-16c_0}}
           & ~~\text{for~solution}~(\ref{DBbrane2})
             ~\text{with}~c_0\rightarrow1/16
             \end{array}\right.  ,
\end{eqnarray}
It is clear that the single brane is localized at $z=0$,
while the two sub-branes are
localized at $z=\pm \delta/2$ and the thickness of the double
brane is $\delta$. In what follows, we mainly discuss the solution
(\ref{DBbrane1}) with the case $c_0\rightarrow-2$ (the double brane case);
the discussion and result for (\ref{DBbrane2}) are similar.
We define a new constant $u$ for convenience as
\begin{eqnarray}
 u=\sqrt{c_0^2-4}.
\end{eqnarray}
Then the double brane case corresponds to $u\rightarrow 0$ now, and we can
further let
\begin{eqnarray}
 u=4~e^{-\delta_0}
\end{eqnarray}
with $\delta_0\gg 1$. The thickness of the double brane for solution
(\ref{DBbrane1}) is
\begin{eqnarray}
 \delta\approx \frac{1}{b v} \ln\frac{4}{u}
       = \frac{\delta_0}{b v}.
\end{eqnarray}
Note that this thickness can be adjusted easily by $\delta_0$,
an integral parameter independent of the scalar potential $V(\phi,\chi)$.
Furthermore, the thickness is independent of the asymmetric factor $\beta$.

\section{The zero mode and resonances of gravity on the Bloch branes}
\label{SecGravityLocalize}

Stability and zero mode of gravity on the symmetric Bloch brane have been analyzed in Ref. \cite{DegenerateBlochbranes}.
Here we further analyze the zero mode and resonances of gravity on the
symmetric and asymmetric Bloch branes corresponding to the
solution (\ref{DBbrane1}), which is rewritten as follows:
\begin{subequations}\label{DBbrane}
\begin{eqnarray}
 {\phi}(y)
   &=&  \frac{u v\sinh (2bvy)}
              {u\cosh (2bvy)-c_{0}},
              \label{DBbranephi} \\
 {\chi}(y)
   &=& \frac{2v}{u\cosh (2bvy)-c_{0}},
              \label{DBbranechi} \\
 e^{2A(y)}
  &=&\left( \frac{c_{0}-u}
         {c_{0}-u\cosh (2bvy)}\right)
          ^{4v^{2}/9}  \exp{(-2\beta y)}  \nonumber \\
  &{\times}&
     \exp{ \left[\frac{4uv^2}{9}
     \left( \frac{u-c_0 \cosh (2bvy)}
                 {(c_0-u \cosh (2bvy))^2}
           -\frac{1}{u-c_0}
     \right)
     \right] } .\label{DBbraneE2A}
\end{eqnarray}\end{subequations}
We will analyze the spectra
of gravity on the thick brane by presenting the potential of the
corresponding Schr\"{o}dinger-like equation of the gravitational KK modes.
In order to obtain the corresponding
mass-independent potential, we perform the coordinate
transformation
\begin{equation}
dz=e^{-A(y)}dy \label{coordinateTransformation}
\end{equation}
to get a conformally flat metric
\begin{equation}
ds^{2}=e^{2A(z)}(\eta_{\mu\nu}dx^{\mu}dx^{\nu}+dz^{2}).
\label{conformallyFlatMetric}
\end{equation}
The analyzing of a full set of fluctuations
of the metric around the background is a complex work. However, the problem
can be simplified when one only considers the transverse and traceless part of the metric
fluctuation \cite{dewolfe}. So,
we consider the following tensor perturbation of the metric:
\begin{eqnarray}
 ds^2&=&\left(e^{2A(z)}\eta_{\mu \nu}
      +\hat{h}_{\mu \nu}(x,z)\right)dx^\mu dx^\nu
      +e^{2A(z)}dz^2\nonumber \\
 &=&e^{2A(z)}\left[\left(\eta_{\mu \nu}
      +h_{\mu \nu}(x,z)\right)dx^\mu dx^\nu
      +dz^2\right].
\label{pertubation_metric}
\end{eqnarray}
Here $ h_{\mu \nu}$ is the tensor perturbation of the metric, and it
satisfies the transverse traceless condition \cite{dewolfe}:
${h_\mu}^\mu=\partial^\nu h_{\mu \nu}=0$.
The equation for $h_{\mu\nu}$ is given by
\begin{equation}
  \partial_{z}^2 h_{\mu \nu}+3 (\partial_{z}A)(\partial_{z}h_{\mu \nu})
      + \eta^{\lambda\rho}\partial_\lambda\partial_\rho h_{\mu \nu} =0.
      \label{Eqs_of_h_mu_nu}
\end{equation}
By performing the following decomposition
\begin{equation}
  h_{\mu \nu}(x,z) = e^{-\frac{3}{2}A}e^{ikx}\varepsilon_{\mu\nu}h(z),
\label{decompsoseh_mu_nu}
\end{equation}
where $k^2=-m^2$ with $m$ the four-dimensional mass of a gravitational KK excitation,
Eq. (\ref{Eqs_of_h_mu_nu}) can be
recast into a Schr\"{o}dinger-like equation
\begin{equation}
      \big(-\partial_{z}^2+V_G(z)\big) h(z)=m^2h(z)
\label{KK_G_Schrodinger_Eqs}
\end{equation}
with the effective potential given by
\begin{eqnarray}
 V_G(z)=\frac{3}{2}\partial_{z}^2 A(z)
        +\frac{9}{4} \big(\partial_{z}A(z))^2.\label{KK_G_potential}
\end{eqnarray}

\subsection{The potential and the zero mode}
\label{SecGravityZeroMode}

\begin{figure}[htb]
\begin{center}
\subfigure[$\beta=0$]{\label{fig_Zeromode_VGy_sDB1}
\includegraphics[width=7cm,height=4.5cm]{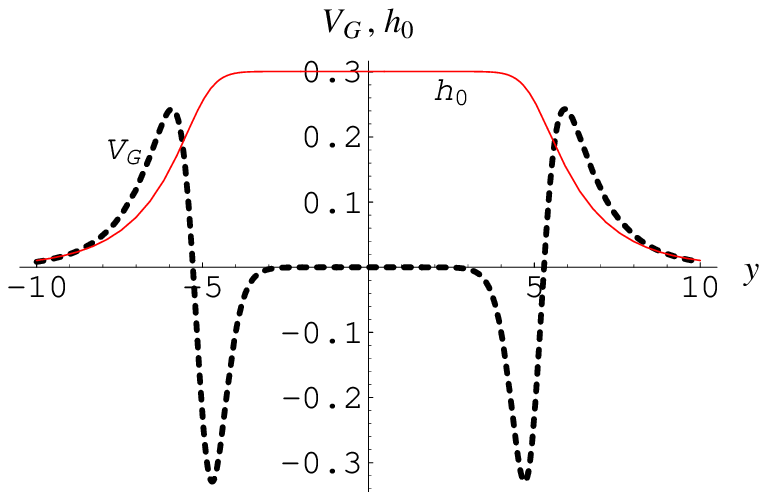}} 
\subfigure[$\beta=0$]{\label{fig_Zeromode_VGz_sDB1}
\includegraphics[width=7cm,height=4.5cm]{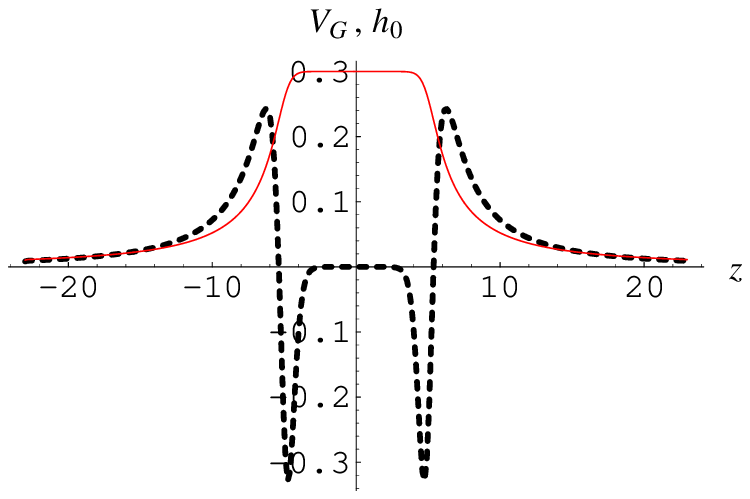}}
\subfigure[$\beta=1/16$]{\label{fig_Zeromode_VGy_aDB1}
\includegraphics[width=7cm,height=4.5cm]{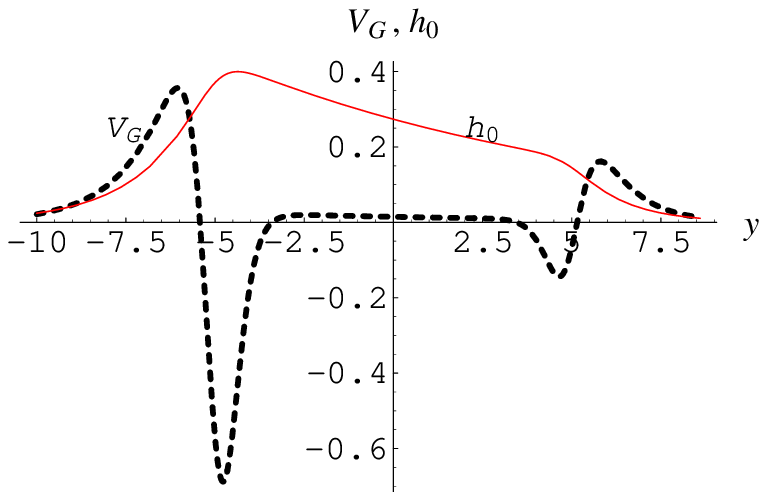}}
\subfigure[$\beta=1/16$]{\label{fig_Zeromode_VGz_aDB1}
\includegraphics[width=7cm,height=4.5cm]{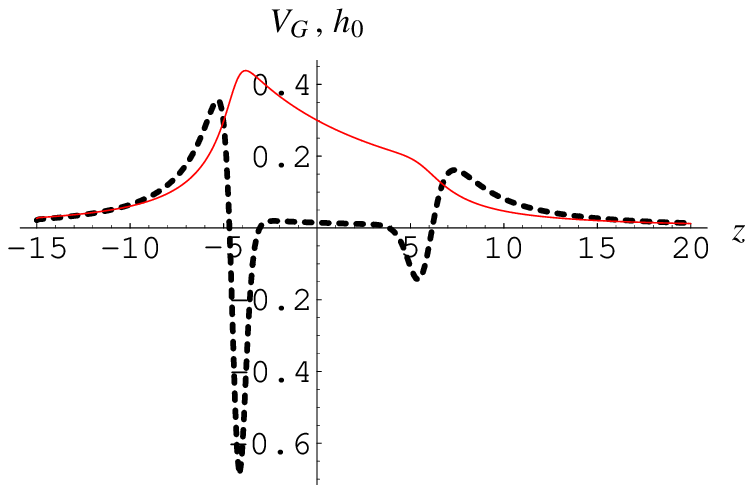}}
\end{center}
\vskip -4mm
 \caption{The shape of the potential $V_G$ (thick dashed lines)
 and the zero mode $h_0$ (red thin lines) for the symmetric
 (up) and asymmetric (down)
 degenerate Bloch branes with $a=b$ in $y$ (left) and
 $z$ (right) coordinates.
 The parameters are set to $b=v=1$, $c_0=-2- 10^{-8}~(\delta_0=9.9)$.}
 \label{fig_ZeroMode_VG_DB1}
\end{figure}

Here, we face the difficulty that the function $y(z)$ cannot be expressed
in an explicit form for the brane solutions given in the previous section.
But we can write the potential $V_G$ as a function of $y$:
\begin{eqnarray}
  V_G(z(y))
     &=& e^{2A(y)}\bigg(\frac{3}{2}\partial_{y}^2 A(y)
          +\frac{15}{4} \big(\partial_{y}A(y)\big)^2\bigg).
\end{eqnarray}
For the asymmetric brane solution corresponding to (\ref{DBbrane}),
the explicit expression of $V_G(z(y))$ is
\begin{widetext}
\begin{eqnarray}
 V_G(z(y))
     &=& \left(\frac{c_0-u}{c_0-u {\cosh}(2bvy)}\right)^{\frac{4 v^2}{9}}
     \exp\left[\frac{4}{9}u v^2
               \left(\frac{u-c_0\cosh(2 b v y)}
                     {\left(c_0-u\cosh(2 b v y)\right)^2}
                     -\frac{1}{u-c_0}
               \right)
               -2{\beta}y  \right]
         \nonumber \\
     & \times & \left\{\frac{5}{108} \left[2 b u^2 v^3  {\sinh}(2bvy)
              \frac{\left(u (5+{\cosh}(4bvy))-6 c_0 {\cosh}(2bvy)\right)}
                   {\left(u{\cosh}(2bvy)-c_0\right)^3}
              +9 \beta \right]^2\right.\nonumber \\
     && ~~~+ \left.2 b^2 u^2 v^4
          \frac{\left(4 u c_0{\cosh}(2bvy)
                 -\left(c_0^2+4\right){\cosh}(4bvy)-3 u^2\right)}
               {\left(u {\cosh}(2bvy)-c_0\right)^4}
        \right\}. \label{V_Gzy_DB1}
\end{eqnarray}
\end{widetext}
The values of $V_G$ at $y=0$ and $y\rightarrow\pm\infty$ are
\begin{eqnarray}
 &&V_G(0)=\frac{15 \beta ^2}{4}
       -\frac{8 b^2 v^4 u^2(2+u^2+u\sqrt{u^2+4})}
             {(u+\sqrt{u^2+4})^4}, \\
 &&V_G(y\rightarrow\pm\infty)\rightarrow0.
\end{eqnarray}
The corresponding zero mode solution is
\begin{eqnarray}
 h_0(z(y))&\propto& e^{\frac{3}{2}A(z(y))}
    = h^S_0(z(y))\; e^{-\frac{3}{2}\beta y},
\end{eqnarray}
where $h^S_0(z(y))$ is the zero mode solution
for the symmetric brane case and is given by
\begin{eqnarray}
 h^S_0(z(y))&\propto& 
   \left(\frac{c_0-u}{c_0-u {\cosh}(2bvy)}\right)^{{v^2}/{3}}
   \nonumber \\
    &\times&  \exp\left\{\frac{u v^2(u-c_0{\cosh}(2bvy))}
                  {3\left(c_0-u {\cosh}(2bvy)\right)^2}
           \right\}.
\end{eqnarray}
For small $u=\sqrt{c_0^2-4}\ll 1$, the brane has two sub-branes
(see Fig. \ref{fig_EnergyDensity_DB1})
and the potential $V_G$ has two subwells located at $y=\pm\delta/2$
(see Fig. \ref{fig_ZeroMode_VG_DB1}).
The zero mode $h_0(z(y))$ is a constant between
the two sub-branes for the symmetric brane
and it is localized on the left sub-brane for the asymmetric brane.
The zero mode represents the four-dimensional graviton; it is also the lowest energy eigenfunction (ground state)
of the Schr\"{o}dinger-like equation (\ref{KK_G_Schrodinger_Eqs})
since it has no zeros. Since the ground state has the lowest
mass square $m_0^2=0$, there is no tachyonic gravitational KK mode.

\subsection{The massive KK modes and the resonances of gravity on the symmetric Bloch brane}
\label{SecGravityMassiveMode}

Since $V_G(z)\rightarrow 0$ when $z\rightarrow\pm\infty$, the
potential provides no mass gap to separate
the gravitational zero mode from the excited KK modes; i.e.,
there exists a continuous gapless spectrum of the KK modes.
The massive modes will propagate along the extra dimension
and those with lower energy will experience an attenuation
due to the presence of the
potential barriers near the location of the brane.

The shape of the potential is strongly dependent on the constant $u$ (or $c_0$).
When $u\rightarrow 0$, two subwells around the corresponding two
sub-brane locations would appear, which could be related to
gravitational resonances. In Refs. \cite{GremmPLB2000,0801.0801}, the authors
found that there exist gravitational resonances on the thick
domain wall, and the resonances could remarkably affect the
modifications of the four-dimensional Newton's law at distances \cite{0801.0801}.
In Refs. \cite{0901.3543,Liu0904.1785,Liu0907.0910},
a similar potential and resonances for left- and
right-handed fermions were also found on
thick branes with and without internal structure.
In what follows, we investigate the massive modes of
gravity by solving numerically Eq. (\ref{KK_G_Schrodinger_Eqs}) with
potential in (\ref{V_Gzy_DB1}). We follow
the method presented in Ref. \cite{Liu0904.1785} to calculate
the probability for finding the massive modes on the Bloch brane.

We impose two kinds of initial conditions in order to
obtain the solutions of the KK modes $h(z)$ from the
second-order differential equations (\ref{KK_G_Schrodinger_Eqs}):
\begin{equation}
 h(0)=h_0, h'(0)=0,   \label{initialCondition1}
\end{equation}
and
\begin{equation}
 h(0)=0, ~~h'(0)=h_1.   \label{initialCondition2}
\end{equation}
The first and second conditions would result in even and odd KK
modes for symmetric potential, respectively.
The constants $h_0$ and $h_1$ for unbound
massive KK modes are arbitrary. The massive KK modes would encounter the tunneling process
across the potential barriers near the brane. And the modes with
different masses would have different lifetimes.

One can interpret $|h(z)|^2$ as the probability for finding the
massive KK modes at the position $z$ along extra dimension
\cite{0901.3543}. According to Refs. \cite{Liu0904.1785,Liu0907.0910},
large relative probabilities for finding massive KK modes
within a narrow range $-z_b<z<z_b$ around the brane location,
called $P_{G}$, would indicate the existence of resonances.
The relative probabilities could be defined in a box with
borders $|z|=10z_{b}$ as follows \cite{Liu0904.1785}:
\begin{equation}
 P_{G}(m)=\frac{\int_{-z_b}^{z_b} |h(z)|^2 dz}
                 {\int_{-10z_b}^{10z_b} |h(z)|^2 dz}.
 \label{Probability}
\end{equation}
Note that the KK mode with mass square $m^2$ much larger than the
maximum of the corresponding potential $V_{G}^{max}$, i.e.,
$m^2 \gg V_{G}^{max}$,
can be approximated as a plane wave mode
$h(z)\propto\cos mz$ or $\sin mz$, and the corresponding
probability would trend to 0.1.

\begin{figure}[htb]
\begin{center}
\includegraphics[width=7cm,height=4.5cm]{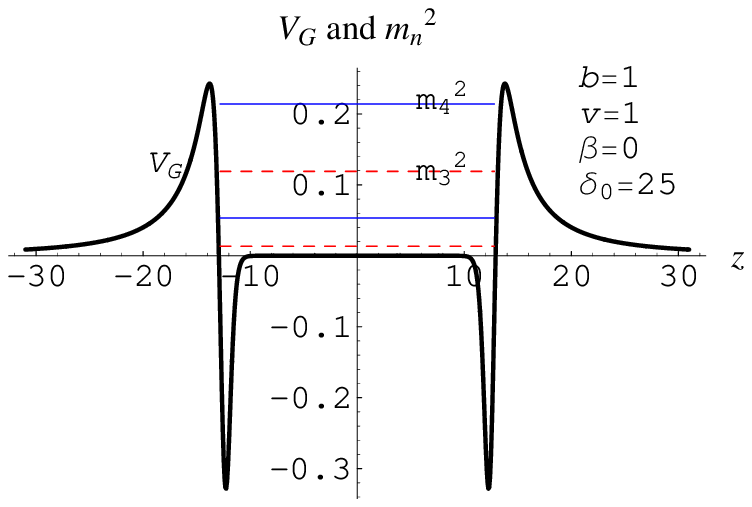}
\includegraphics[width=7cm,height=4.5cm]{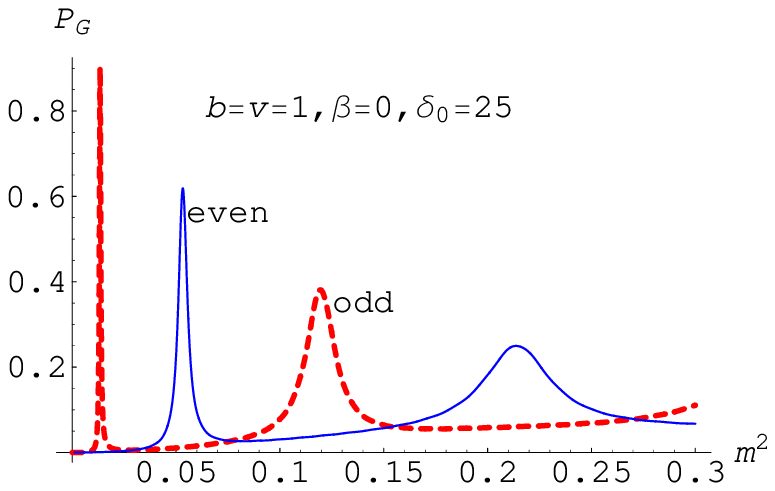}
\includegraphics[width=7cm,height=4.5cm]{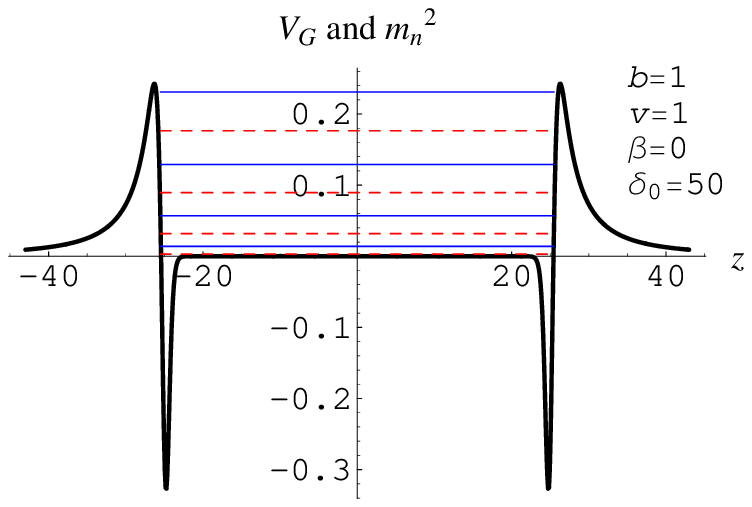}
\includegraphics[width=7cm,height=4.5cm]{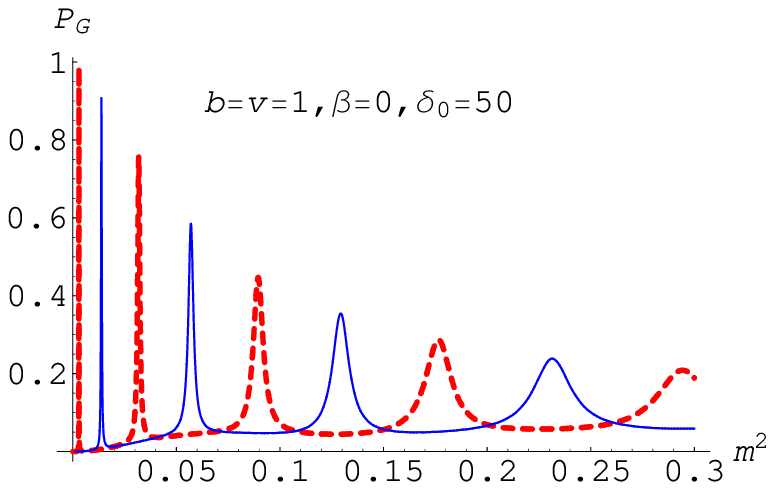}
\includegraphics[width=7cm,height=4.5cm]{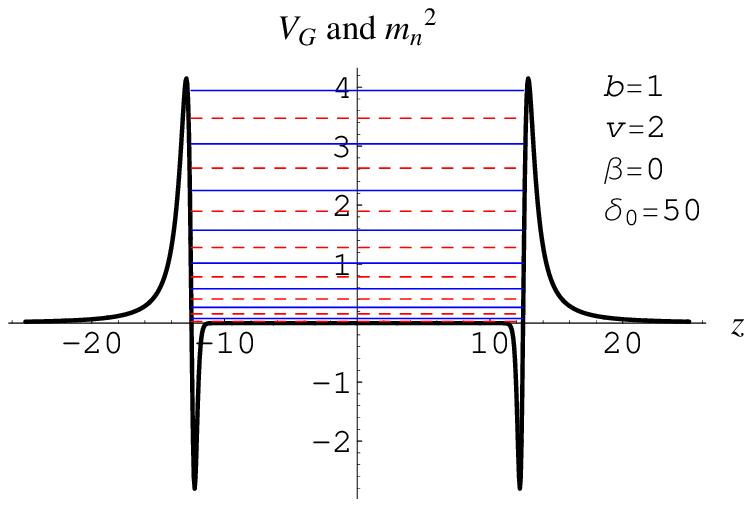}
\includegraphics[width=7cm,height=4.5cm]{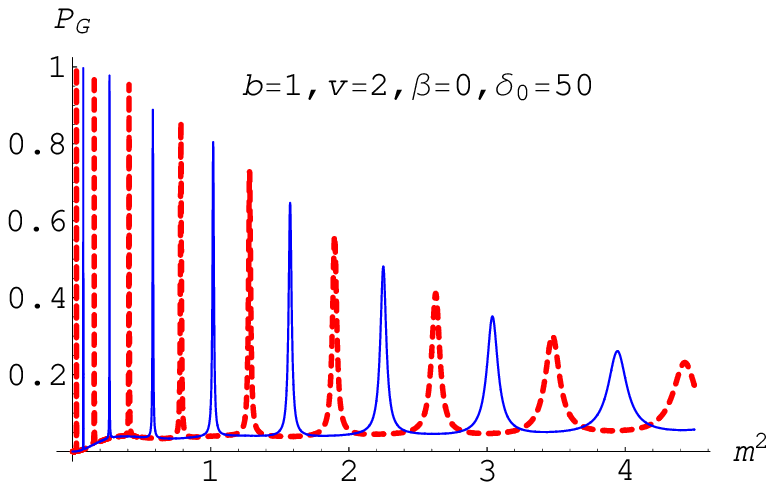}
\end{center}\vskip -4mm
 \caption{The potential $V_G(z)$, the resonance spectrum $m_n^2$ and
 the probability $P_G(m^2)$ of the gravitational KK modes $h(z)$.
 In the figures of $V_G(z)$ and $m_n^2$,
 $V_G(z)$ is denoted by black thick lines,
 $m_n^2$ for the odd resonance KK mode is denoted by red dashed lines,
 and $m_n^2$ for the even resonance KK mode is denoted by blue thin lines.
 In the figures of $P_G(m^2)$, the curves of $P_G(m^2)$ for odd and even
 modes $h(z)$ are denoted by red dashed lines and blue thin lines, respectively.
 }
 \label{fig_resonance_graviton}
\end{figure}

We investigate the massive KK modes of gravity by solving numerically Eq.
(\ref{KK_G_Schrodinger_Eqs}). For the set
of parameters $a=b=1, v=1,~\beta=0$, and $\delta_0=25~($i.e.,
$c_0=-2-7.71\times 10^{-22})$, we find four peaks located at
$m^2=\{0.0134$, $0.0533$, $0.1195$, $0.2137\}$ for the KK modes of gravity (see Fig. \ref{fig_resonance_graviton}).
These peaks are related with resonances of gravity,
which are long-lived massive gravitational excitations on the brane.
Except several peaks, the curves grow at first, and then
stably trend to $z_b/z_{max}=0.1$. The reason is that KK modes
with small $m^2 (\ll V_{G}^{max})$ will be damped near the brane
and oscillate away from the brane, while those modes with large $m^2
(\gg V_{G}^{max})$ can be approximated as plane wave modes
$h(z)\propto\cos mz$ or $\sin mz$.
Note that any gravitational excitation with $m^2>0$
produced on the brane cannot correspond exactly to a single
resonant KK mode because it has a wave function truly localized on
the brane \cite{mass5-Dscalar}. It is a wave packet composed of the
continuum modes with a Fourier spectrum peaked around one of the
resonances.

\begin{figure}[htb]
\begin{center}
\includegraphics[width=7cm,height=4.5cm]{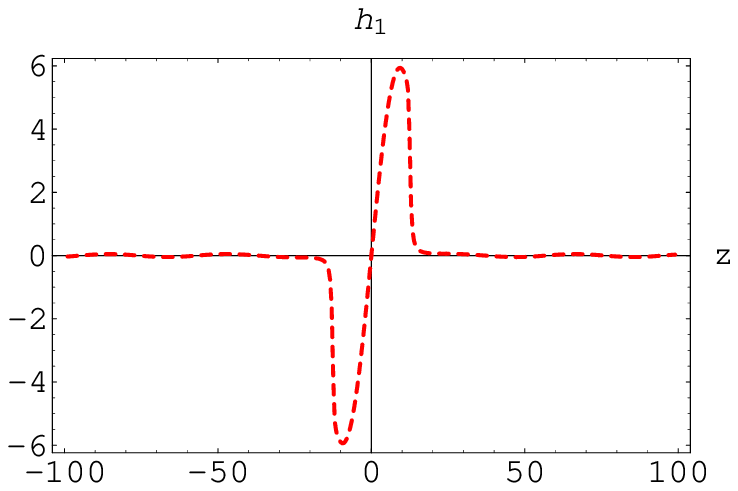}
\includegraphics[width=7cm,height=4.5cm]{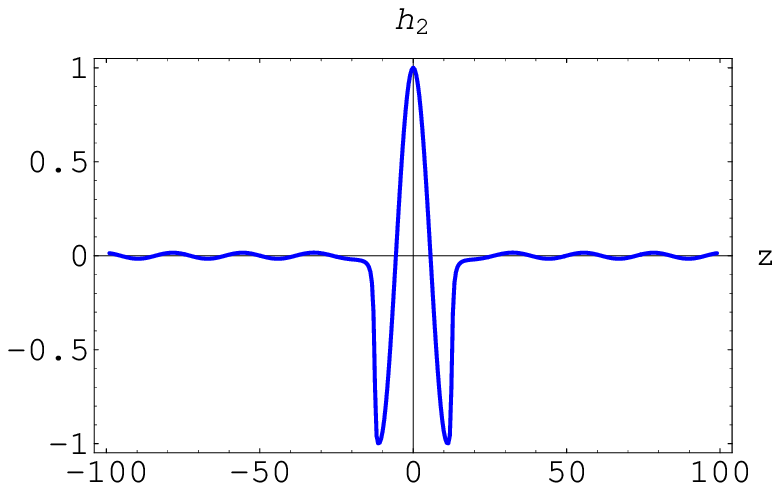}
\includegraphics[width=7cm,height=4.5cm]{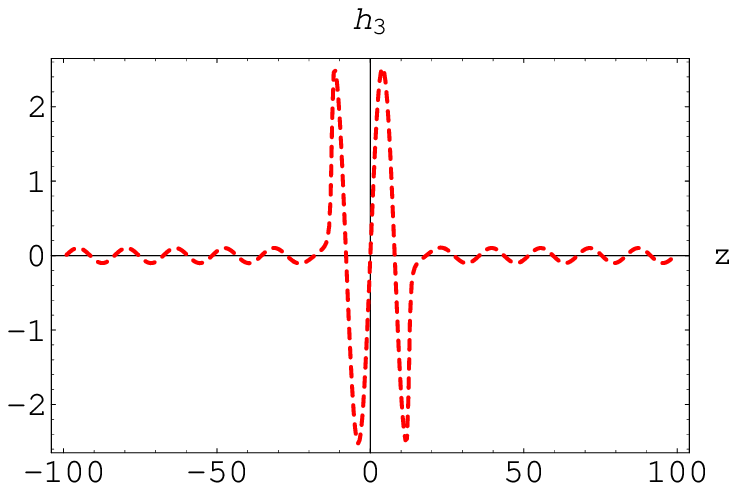}
\includegraphics[width=7cm,height=4.5cm]{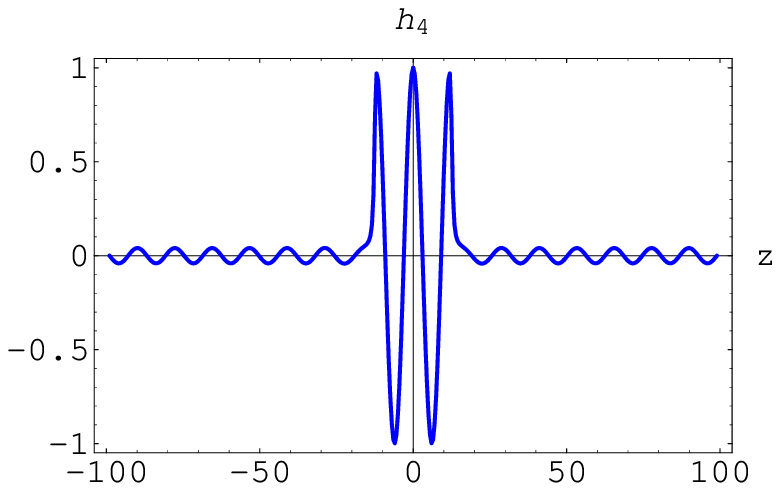}
\includegraphics[width=7cm,height=4.5cm]{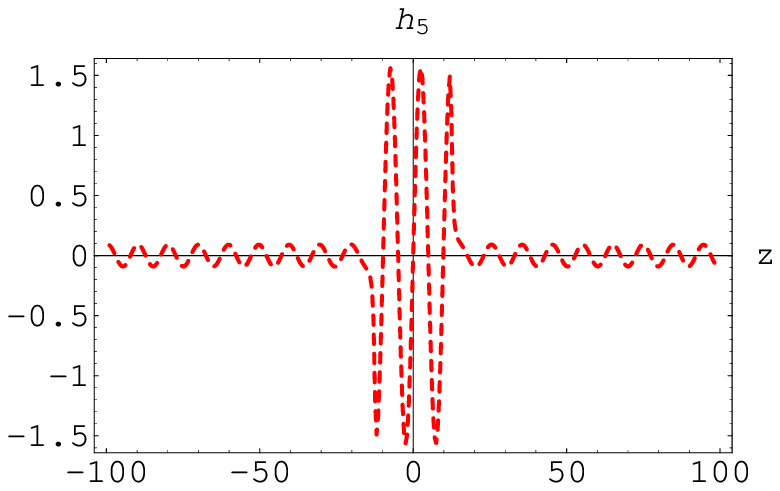}
\includegraphics[width=7cm,height=4.5cm]{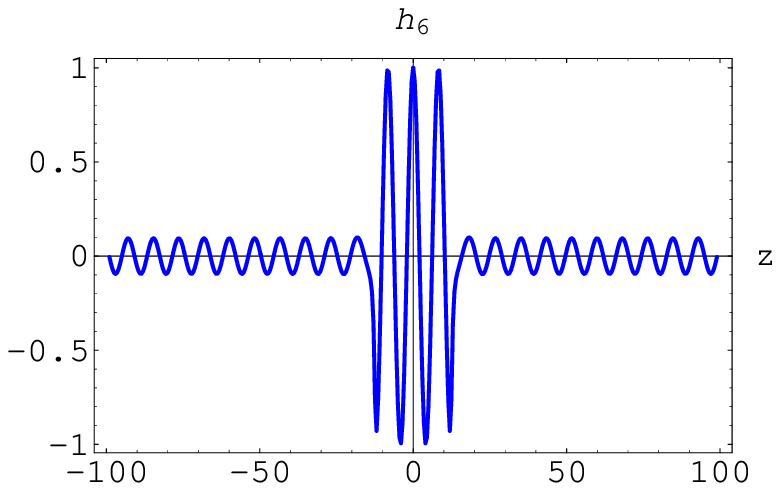}
\end{center}\vskip -4mm
 \caption{The lower level ($n=~1-6$) resonance KK modes $h_n(z)$ of gravity
 with the parameters set as $a=b=1, v=2,~\beta=0$ and $\delta_0=50$.
 The odd and even resonance KK modes are denoted by the red dashed lines
 and the blue continuous lines, respectively.}
 \label{fig_resonance_gravitonKKmodes}
\end{figure}

For $a=b=1, v=1,~\beta=0$, and $\delta_0=50$,
i.e., a brane with double width compared to the one with the same values of
$a,~b,~v,~\beta$ but $\delta_0=25$, we find eight resonances for gravity,
and the mass spectrum of the resonances is calculated as
\begin{eqnarray}
  m^2 &=&\{0.0029, 0.0138, 0.0318, 0.0570,0.0895, \nonumber \\
      && ~ 0.1293, 0.1765, 0.2313\}.
\end{eqnarray}
For $a=b=1, v=2,~\beta=0$, and $\delta_0=50$,
i.e., a brane with the same width compared to the one with
$a=b=1, v=1,~\beta=0$, and $\delta_0=50$, we find sixteen resonances
with the mass spectrum given by
\begin{eqnarray}
  m^2 &=&\{0.0283, 0.0777, 0.1572, 0.2676, 0.4091, 0.5815, \nonumber \\
      && ~ 0.7850,1.0180,1.2814, 1.5748,1.8968,2.2492, \nonumber \\
      && ~ 2.6283, 3.0389,3.4731,3.9435\}.
\end{eqnarray}
We can see that the number of the resonances increases with
the width of the double brane and the value of $v$,
which is related with the VEV of the scalar $\phi$.
The potential $V_G(z)$, the resonance spectra $m_n^2$ and
the probability $P_G(m^2)$ of the gravitational KK mode $h_n(z)$ are shown
in Fig. \ref{fig_resonance_graviton}.
Figure. \ref{fig_resonance_gravitonKKmodes} shows the lower level resonance
KK modes of gravity. The $n=0$ level mode is in fact the only one
bound state, namely, the four-dimensional massless graviton $h_0$.

\subsection{{The resonances of gravity on the asymmetric Bloch brane}}

{In this subsection, we will calculate the resonances of gravity on the asymmetric Bloch brane. Now the effective potential is asymmetric, so the corresponding KK modes have no definite parity anymore. Therefore, the relative probability method used in the last subsection is not convenient for the asymmetric case here. We will adopt the transfer matrix method first introduced in Ref.~\cite{Landim2011} and subsequently used in Refs.~\cite{Landim2012,Alencar2012,zhao2013}. In order to check the consistency of the two methods, we will first calculate the resonances of gravity on the symmetric Bloch brane with the transfer matrix method. Then we will apply it to the asymmetric case.

In this method, the interval of extra dimension, outside of which the potential is small enough, is divided into $N$ intervals with enough large $N$, and the coordinate is denoted by $z_0,z_1,\cdots,z_N$. In each interval the potential is replaced by a square potential barrier $V_{i}$ and the Schr\"{o}dinger-like equation is solved. Then with the continuity of the KK modes and their first derivatives at each point $z_i$ and the iterative method, one can lastly get the transmission coefficient $T$ \cite{zhao2013}:
\begin{equation}
T=\frac{k_1}{k_N}\frac{1}{\mid M_{22}\mid^2}
 =\sqrt{\frac{\mid V_1-E_n\mid}{\mid V_N-E_n\mid}},\label{T}
\end{equation}
where $E_n=m_n^2$ is the ``energy" of the particle, and $M$ is the product of the $N$ transfer coefficient matrixes $M_i$ with dimension $2\times 2$. Note that the transmission coefficient $T$ (\ref{T}) for a symmetric potential (for which $V_1=V_N$ for a symmetric coordinate interval) will reduce to the one given in Ref.~\cite{Landim2011}: $T={1}/{\mid M_{22}\mid^2}$.

For a given potential, the transmission coefficient $T$ is dependent on the value of the mass square $m^2$. So it can be considered as a function of $m^2$, like the case of $P(m^2)$. If $T(m^2)$ has a peak at some $m^2$, then we will have a higher probability to find this massive KK mode on the brane, and we will call this KK mode a resonant state. This is coincident with the relative probability method, which can be seen from Figs.~\ref{fig_resonance_graviton} and ~\ref{fig_resonance_transmissivity_sym_gravity}. In Fig.~\ref{fig_resonance_transmissivity_sym_gravity}, each peak corresponds to a resonant KK state. For the symmetric potential case, the transmission coefficient for each resonance is almost 1. Those resonances with smaller mass square have sharper resonance peaks, which is similar to the results of the relative probability method obtained in the last subsection.

The asymmetric case is shown in Fig.~\ref{fig_resonance_transmissivity_Asym_gravity}. We note here that the resonant KK modes should only correspond to those peaks with mass square smaller than the small maximum of the asymmetric potential, which is respectively 0.2, 0.16, and 0.1 for the three potentials plotted in Fig.~\ref{fig_resonance_transmissivity_Asym_gravity}, because for massive KK modes, there is a ``quasipotential well" that can trap these KK modes at a finite time. The lifetime can also be estimated from the half-width of the peak. The numerical results show that the number and the transmission coefficients of the resonant KK modes will decrease with the asymmetric factor $\beta$. For larger enough asymmetric factor, there will no quasipotential well and hence no resonance anymore, even though there is still a resonance structure in the $T-m^2$ picture. On the other hand, we can see that for a potential, the resonance with smaller mass square has a longer lifetime, which is the same as the symmetric brane case.

\begin{figure}[htb]
\begin{center}
\includegraphics[width=7cm,height=4.5cm]{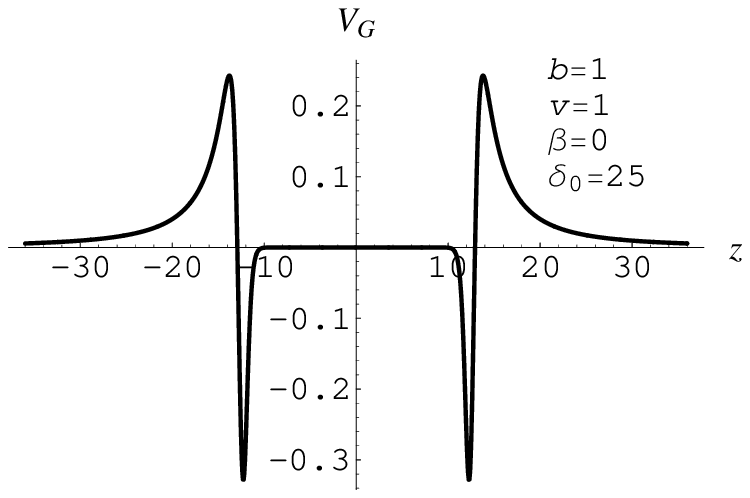}
\includegraphics[width=7cm,height=4.5cm]{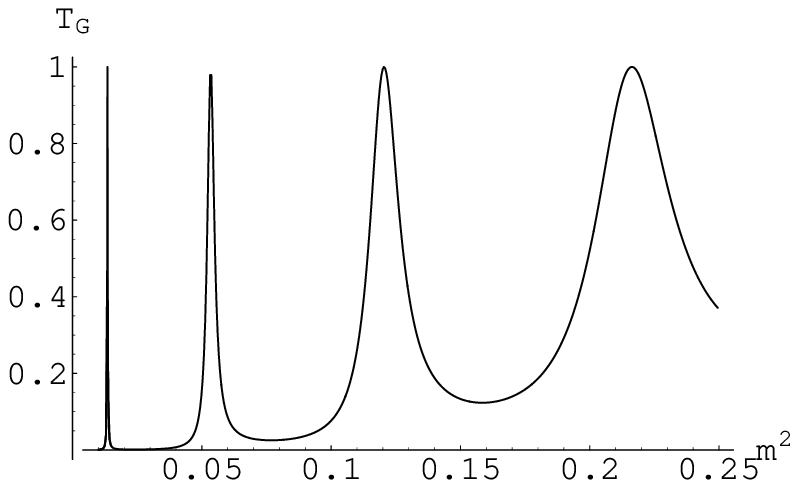}
\includegraphics[width=7cm,height=4.5cm]{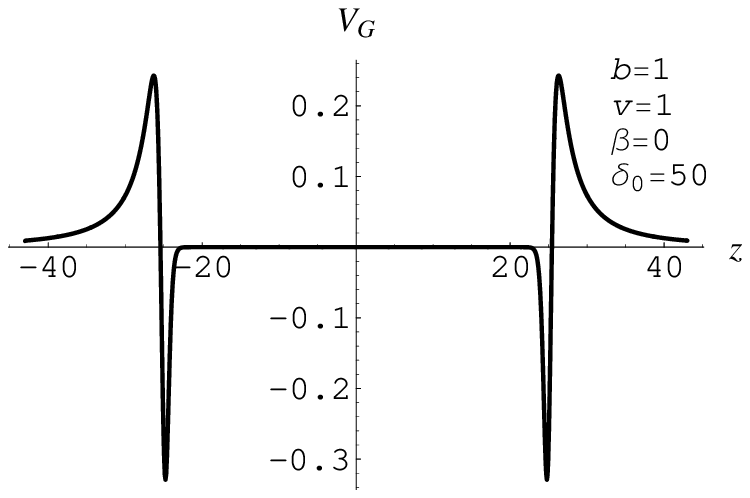}
\includegraphics[width=7cm,height=4.5cm]{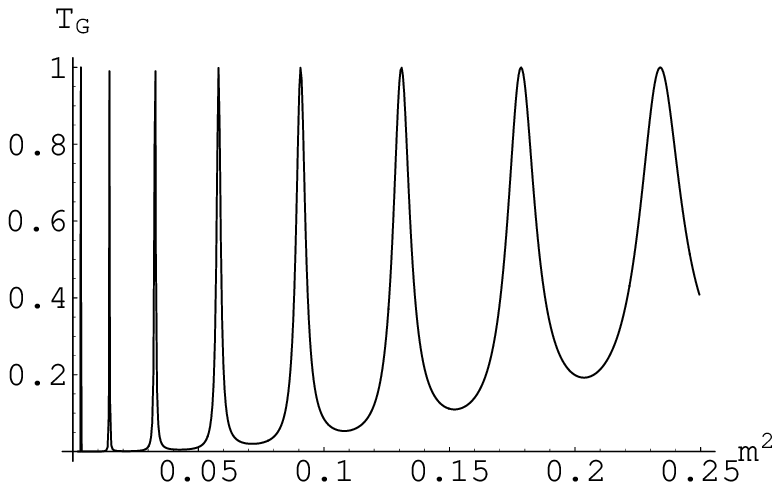}
\includegraphics[width=7cm,height=4.5cm]{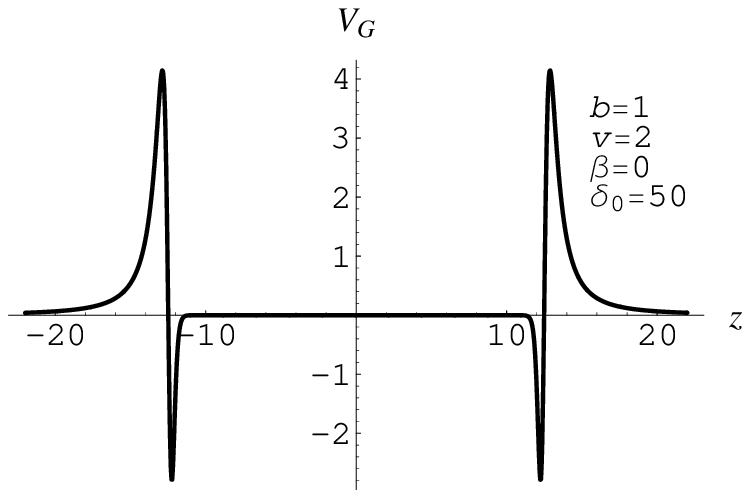}
\includegraphics[width=7cm,height=4.5cm]{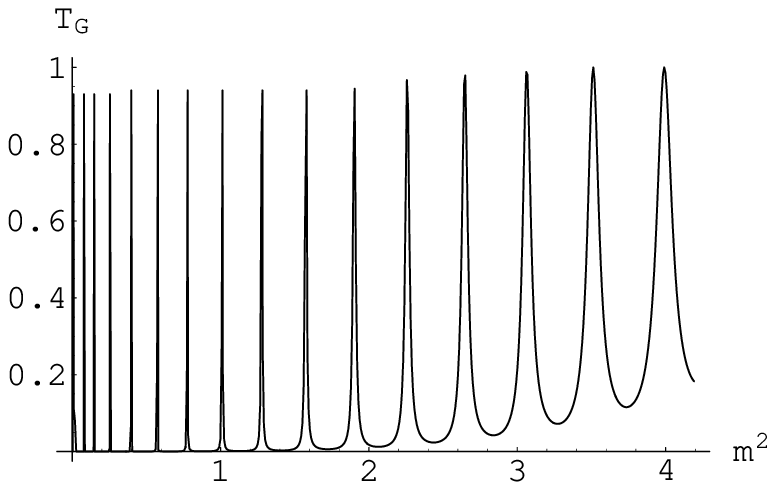}
\end{center}\vskip -4mm
 \caption{The potential $V_G(z)$ and the transmission coefficient $T_G$ of gravity KK modes on the symmetric Bloch brane.}
 \label{fig_resonance_transmissivity_sym_gravity}
\end{figure}

\begin{figure}[htb]
\begin{center}
\includegraphics[width=7cm,height=4.5cm]{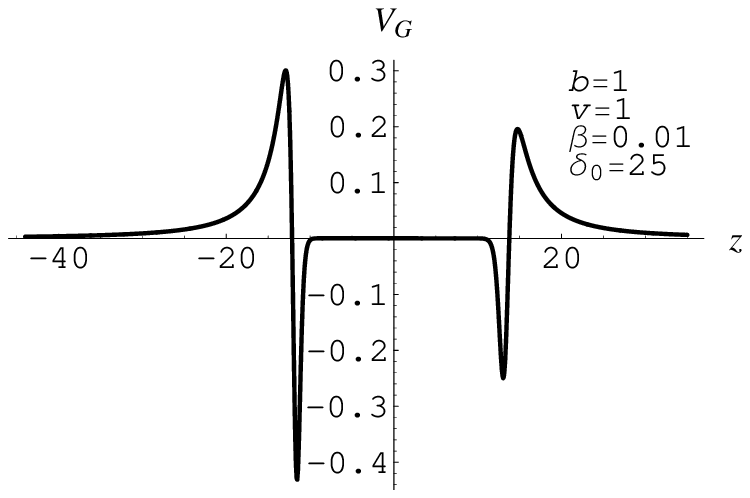}
\includegraphics[width=7cm,height=4.5cm]{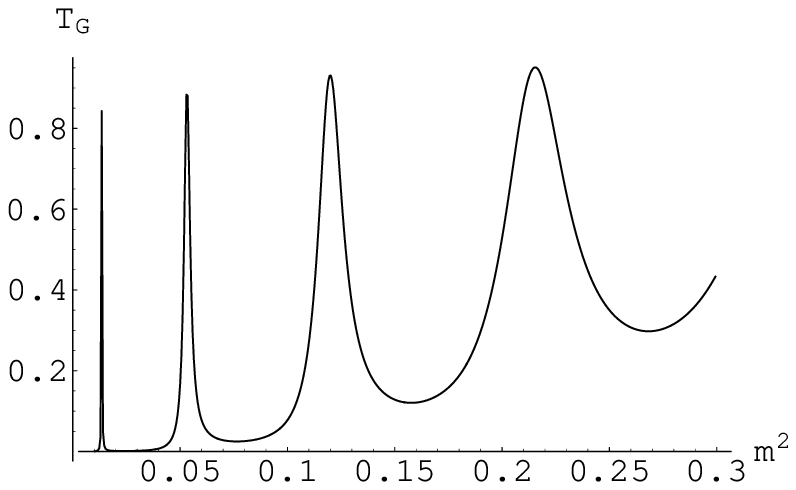}
\includegraphics[width=7cm,height=4.5cm]{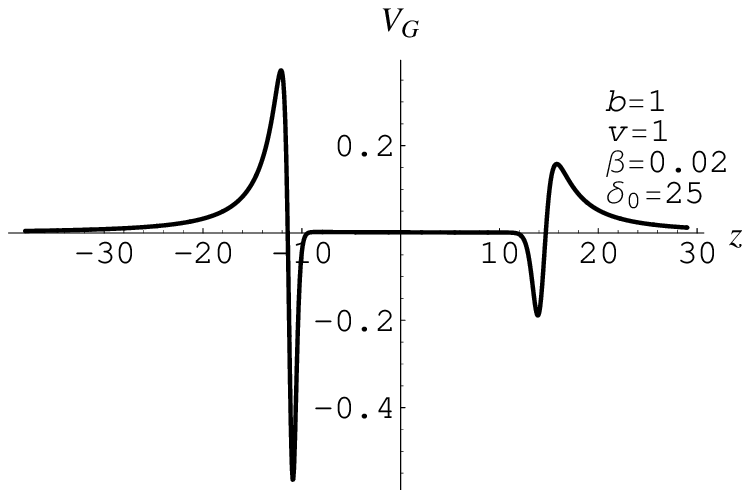}
\includegraphics[width=7cm,height=4.5cm]{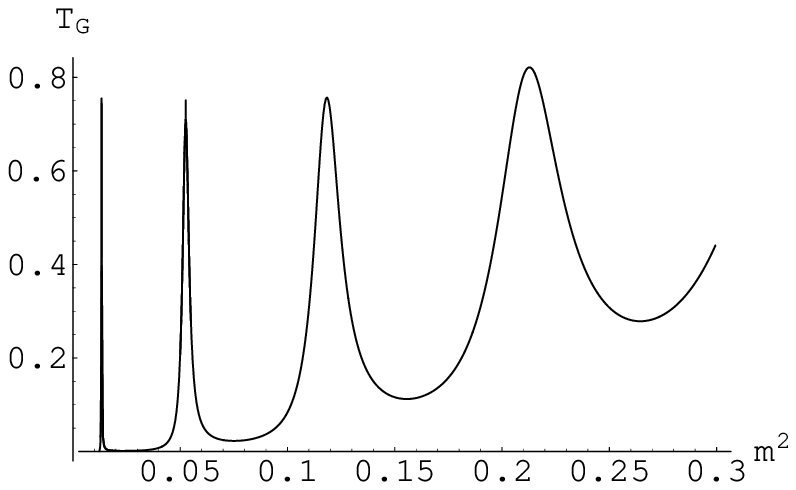}
\includegraphics[width=7cm,height=4.5cm]{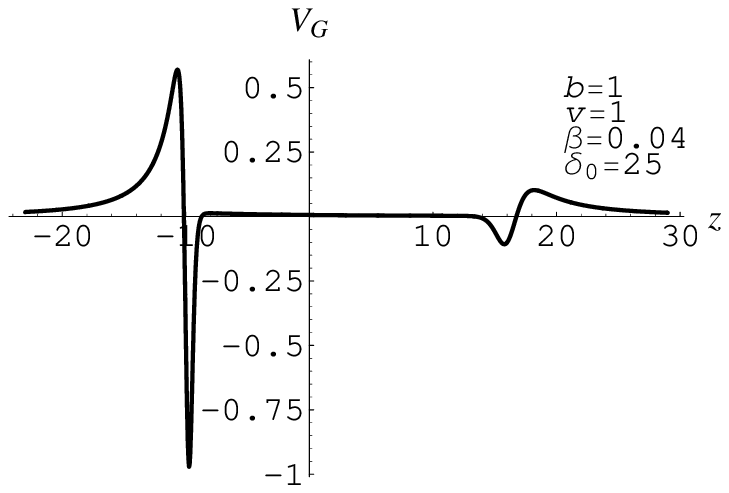}
\includegraphics[width=7cm,height=4.5cm]{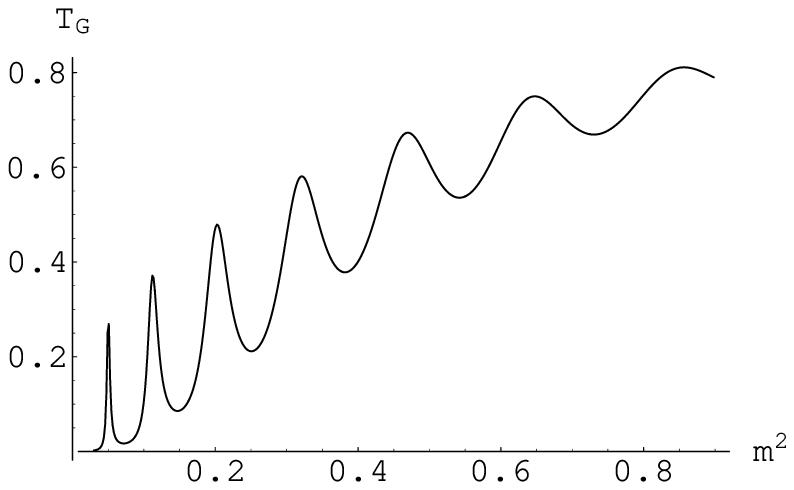}
\end{center}\vskip -4mm
 \caption{The potential $V_G(z)$ and the transmission coefficient $T_G$ of gravity KK modes on the asymmetric Bloch brane.}
 \label{fig_resonance_transmissivity_Asym_gravity}
\end{figure}

\section{Localization and mass spectra of fermions on the Bloch brane}
\label{SecFermionLocalize}

In this section, we would like to investigate the localization problem of fermions on the symmetric and asymmetric thick branes
given in Eqs. (\ref{DBbrane}) by introducing scalar-fermion coupling.
We will analyze the spectra of fermions on the thick brane
{by presenting }the potential of the Schr\"{o}dinger-like equation of the
fermion KK modes.

In five dimensions, fermions are four-component spinors and their
Dirac structure is described by $\Gamma^M= e^M _{\bar{M}}
\Gamma^{\bar{M}}$ with $\{\Gamma^M,\Gamma^N\}=2g^{MN}$, where
$\bar{M}, \bar{N}, \cdots =0,1,2,3,5$ denote the five-dimensional
local Lorentz indices, and $\Gamma^{\bar{M}}$ are the flat gamma
matrices in five dimensions. In our setup,
$\Gamma^M=(\text{e}^{-A}\gamma^{\mu},\text{e}^{-A}\gamma^5)$, where
$\gamma^{\mu}$ and $\gamma^5$ are the usual flat gamma matrices in
the Dirac representation. The Dirac action of a massless spin-1/2
fermion coupled to the scalar is
\begin{eqnarray}
S_{1/2} = \int d^5 x \sqrt{-g} \left(\bar{\Psi} \Gamma^M D_M
\Psi-\eta \bar{\Psi} F(\phi) \Psi\right), \label{DiracAction}
\end{eqnarray}
where the covariant derivative $D_M$ is defined as $D_M\Psi =
(\partial_M + \omega_M) \Psi$ with the spin connection $\omega_M=
\frac{1}{4} \omega_M^{\bar{M} \bar{N}} \Gamma_{\bar{M}}
\Gamma_{\bar{N}}$. With the metric (\ref{linee}), the nonvanishing
components of the spin connection $\omega_M$ are
\begin{eqnarray}
  \omega_\mu =\frac{1}{2}(\partial_{z}A) \gamma_\mu \gamma_5. \label{eq4}
\end{eqnarray}
Then the five-dimensional Dirac equation is read as
\begin{eqnarray}
 \left[ \gamma^{\mu}\partial_{\mu}
         + \gamma^5 \left(\partial_z  +2 \partial_{z} A \right)
         -\eta\; \text{e}^A F(\phi)
 \right ] \Psi =0, \label{DiracEq1}
\end{eqnarray}
where $\gamma^{\mu} \partial_{\mu}$ is the Dirac operator on the
brane. Note that the sign of the coupling $\eta$ between the spinor
$\Psi$ and the scalar $\phi$ is arbitrary and represents a coupling
either to kink or to antikink domain wall. For definiteness, we
will consider in what follows only the case of a kink coupling, and
thus assume that $\eta>0$.

Now we study the above five-dimensional Dirac equation. Because of
the Dirac structure of the fifth gamma matrix $\gamma^5$, we expect
the left- and right-handed projections of the four-dimensional part to
behave differently. From the equation of motion (\ref{DiracEq1}), we
will search for the solutions of the general chiral decomposition
\begin{eqnarray}
 \Psi(x,z) &=& \sum_n \Big(
       \psi_{Ln}(x) \hat{f}_{Ln}(z)
       +\psi_{Rn}(x) \hat{f}_{Rn}(z)
       \Big), \nonumber \\
 &=&  \text{e}^{-2A}\sum_n  \Big(  \psi_{Ln}(x) f_{Ln}(z)
       +\psi_{Rn}(x) f_{Rn}(z) \Big),
\end{eqnarray}
where $\hat{f}_{L,R}(z)=\text{e}^{-2A}{f}_{L,R}(z)$,
$\psi_{Ln}(x)=-\gamma^5 \psi_{Ln}(x)$ and
$\psi_{Rn}(x)=\gamma^5 \psi_{Rn}(x)$ are the left-handed and
right-handed components of a four-dimensional Dirac field,
respectively, the sum over $n$ can be both discrete and continuous.
Here, we assume that $\psi_{Ln}(x)$ and $\psi_{Rn}(x)$ satisfy the
four-dimensional massive Dirac equations
$\gamma^{\mu}\partial_{\mu}\psi_{Ln}(x)=m_n\psi_{R_n}(x)$ and
$\gamma^{\mu}\partial_{\mu}\psi_{Rn}(x)=m_n\psi_{L_n}(x)$. Then
$f_{Ln}(z)$ and $f_{Rn}(z)$ satisfy the following coupled
equations
\begin{subequations}
\begin{eqnarray}
 \left[\partial_z + \eta\;\text{e}^A F(\phi) \right]f_{Ln}(z)
  &=&  +m_n f_{Rn}(z), \label{CoupleEq1a}  \\
 \left[\partial_z- \eta\;\text{e}^A F(\phi) \right]f_{Rn}(z)
  &=&  -m_n f_{Ln}(z). \label{CoupleEq1b}
\end{eqnarray}\label{CoupleEq1}
\end{subequations}
From the above coupled equations, we get the Schr\"{o}dinger-like
equations for the KK modes of the left- and right-handed fermions
\begin{subequations}
\begin{eqnarray}
  \big(-\partial^2_z + V_L(z) \big)f_{Ln}
            &=&\mu_n^2 f_{Ln},
   \label{SchEqLeftFermion}  \\
  \big(-\partial^2_z + V_R(z) \big)f_{Rn}
            &=&\mu_n^2 f_{Rn},
   \label{SchEqRightFermion}
\end{eqnarray}
\end{subequations}
where the effective potentials are given by
\begin{subequations}
\begin{eqnarray}
  V_L(z)&=& \big(\eta\;\text{e}^{A}   F(\phi)\big)^2
     - \partial_z\big(\eta\;\text{e}^{A}   F(\phi)\big), \label{VL}\\
  V_R(z)&=&   V_L(z)|_{\eta \rightarrow -\eta}. \label{VR}
\end{eqnarray}\label{Vfermion}
\end{subequations}
In order to obtain the standard four-dimensional action for the massive
chiral fermions:
\begin{eqnarray}
 S_{1/2} &=& \int d^5 x \sqrt{-g} ~\bar{\Psi}
     \left(  \Gamma^M (\partial_M+\omega_M)
     -\eta F(\phi)\right) \Psi  \nn \\
  &=& \sum_{n}\int d^4 x \left(~\bar{\psi}_{Rn}
      \gamma^{\mu}\partial_{\mu}\psi_{Rn}
        -~\bar{\psi}_{Rn}m_{n}\psi_{Ln} \right) \nn \\
  &+&\sum_{n}\int d^4 x \left(~\bar{\psi}_{Ln}
      \gamma^{\mu}\partial_{\mu}\psi_{Ln}
        -~\bar{\psi}_{Ln}m_{n}\psi_{Rn} \right)  \nn \\
  &=&\sum_{n}\int d^4 x
    ~\bar{\psi}_{n}
      (\gamma^{\mu}\partial_{\mu} -m_{n})\psi_{n},
\end{eqnarray}
we need the following orthonormality conditions for $f_{L_{n}}$ and
$f_{R_{n}}$:
\begin{eqnarray}
 &&\int_{-\infty}^{\infty} f_{Lm} f_{Ln}dz=\delta_{mn}
   = \int_{-\infty}^{\infty} f_{Rm} f_{Rn}dz,~~ \nonumber \\
 &&\int_{-\infty}^{\infty} f_{Lm} f_{Rn}dz=0.
 \label{orthonormality}
\end{eqnarray}

It can be seen that, in order to localize the left- or right-handed
fermions, there must be some kind of scalar-fermion coupling, and
the effective potential $V_L(z)$ or $V_R(z)$ should have a minimum
at the location of the brane. Furthermore, for the kink
configuration of the scalar $\phi(z)$ (\ref{DBbranephi}), $F(\phi(z))$
should be an odd function of $\phi(z)$ when one demands that
$V_{L,R}(z)$ are invariant under $Z_2$ reflection symmetry
$z\rightarrow -z$. Thus we have $F(\phi(0))=0$ and
$V_L(0)=-V_R(0)=-\eta\partial_z (e^AF(\phi(0)))$, which results in the
well-known conclusion: only one of the massless left- and right-handed
fermions could be localized on the brane. The spectra are
determined by the behavior of the potentials at infinity. For
$V_{L,R}\rightarrow 0$ as $|z|\rightarrow \infty$, one of the
potentials would have a volcanolike shape and there exists only a
bound massless mode followed by a continuous gapless spectrum of
KK states, while another could not trap any bound states and the
spectrum is also continuous and gapless. The Yukawa
coupling $F(\phi)=\phi$ and the generalized coupling
$F(\phi)=\phi^k$ with positive odd integer $k\;(\geq3)$ belong to
this type. For $V_{L,R}\rightarrow V_{\infty}=$ positive constant as
$|z|\rightarrow \infty$, those modes with $m_n^2<V_{\infty}$
belong to a discrete spectrum and modes with $m_n^2>V_{\infty}$
contribute to a continuous one. If the
potentials increase as $|z|\rightarrow \infty$, the spectrum is
discrete. There are a lot of couplings for this case. The concrete
behavior of the potentials is dependent on the function $F(\phi)$.
In what follows, we will discuss in detail the Yukawa
coupling $F(\phi)=\phi$.

\subsection{The potentials}
\label{sec3.1}

We mainly consider the Yukawa coupling {$\eta\bar{\Psi}\phi\Psi$},
for which the explicit forms of the potentials (\ref{Vfermion}) are
\begin{subequations}
\begin{eqnarray}
 V_L &=& \eta\left( \frac{u-c_{0}}{u\cosh(2bvy)-c_{0}}\right)^{\frac{4v^2}{9}} \text{e}^{-2\beta y }
  \nonumber \\
 &&
  \times\exp\left\{ \frac{4uv^2}{9}
     \left( \frac{u-c_0 \cosh (2bvy)}{(u \cosh (2bvy)-c_0)^2}
           -\frac{1}{u-c_0}   \right)   \right \}
       \nonumber \\
 && \times
    \bigg\{
  \frac{8 b u^3 v^4 [u-c_0\cosh(2bvy)]\text{sinh}^2(2bvy)}
             {9\left(u \cosh(2bvy)-c_0\right)^4}    \nonumber \\
 &&
       +\frac{u^2 v^2 (18b+4bv^2+9\eta) \text{sinh}^2(2bvy)}
             {9\left(u \cosh(2bvy)-c_0\right)^2}\nonumber \\
 &&   +\frac{u v(\beta\text{sinh}(2bvy)-2bv\cosh(2bvy))}
             {\left(u \cosh(2bvy)-c_0\right)} \nonumber \\
 &&   +\frac{4 b u^2 v^4 c_0 \text{sinh}^2(2bvy)}
             {9\left(u \cosh(2bvy)-c_0\right)^3}
      \bigg\},    \label{VL} \\
  V_R &=& V_L|_{\eta \rightarrow -\eta},\label{VSR_CaseI}
\end{eqnarray}
\end{subequations}
for the symmetric (let $\beta=0$) and asymmetric brane solutions.
The double thick brane corresponds to the case $c_0\rightarrow -2$
and $u=4e^{-\delta_0}$ with $\delta_0 \gg 1$.

Both potentials have the asymptotic behavior:
$V_{L,R}(y\rightarrow\pm \infty)\rightarrow0$. The values of
the potentials for left- and right-handed fermions at $y = 0$ are
given by
\begin{equation}
 V_L(0) =-V_R(0)
 = -\frac{2uv^2 b}{(u-c_0)}\eta.
\end{equation}
It is clear that, for a given coupling constant $\eta$, the values of
the potentials for left- and right-handed fermions at $y=0$ are
opposite. The parameter $u$ is positive, $c_0<-2$, and $b$ and $\eta$ can be
positive and negative. Considering that the potentials of left- and
right-handed fermion KK modes are partner potentials,
we will only take the positive values of $b$ and $\eta$
without loss of generality.
We recall that the brane is a single brane and a double brane
for $u\gg 1 ~(|c_0|\gg2)$ and $u\ll 1 ~(c_0\rightarrow -2)$, respectively.
For the double brane case, the potentials at the location of the brane are
\begin{equation}
 V_L(0) =-V_R(0)= -uv^2 b\eta\approx 0.
\end{equation}
The shapes of the potentials are shown in Fig. \ref{fig_VLVR_delta_0} for different values of $u$ or $\delta_0$.
From the figure we see that $V_L(z)$ is a modified volcano
type potential for the single brane scenario and has
a well. While for the double brane case with $u\ll 1$, the
corresponding potential $V_L(z)$ for left-handed fermions has a double well, and the
potential $V_R(z)$ for right-handed fermions has a single well,
which indicates that there may exist resonant (quasilocalized) KK
modes of fermions. Hence, the shape of the potentials is relative to the inner
structure of the brane, or equivalently, it depends partly on the
warp factor $e^{2A}$, and partly on the configuration of the scalar
$\phi$. The effect of other parameters to the potentials is shown in Figs.
\ref{fig_VLVR_eta} and \ref{fig_VLVR_beta}.

On the other hand, we note that $V_L(z)\rightarrow 0$ from above when
$y\rightarrow\pm\infty$, so the potential for left-handed fermions
provides no mass gap to separate the fermion zero mode
from the excited KK modes. For right-handed fermions, the corresponding
potential $V_R(z)\rightarrow 0$ when
$z\rightarrow\pm\infty$, and $V_R(0)> 0$ when
$z=0$, and there is no bound right-handed
fermion zero mode. In fact, this is a simple consequence of the
fact that $V_L$ and $V_R$ are partner potentials.
For both left- and right-handed fermions, there
exists a continuous gapless spectrum of the KK modes.

\begin{figure}[htb]
\begin{center}
\includegraphics[width=7cm]{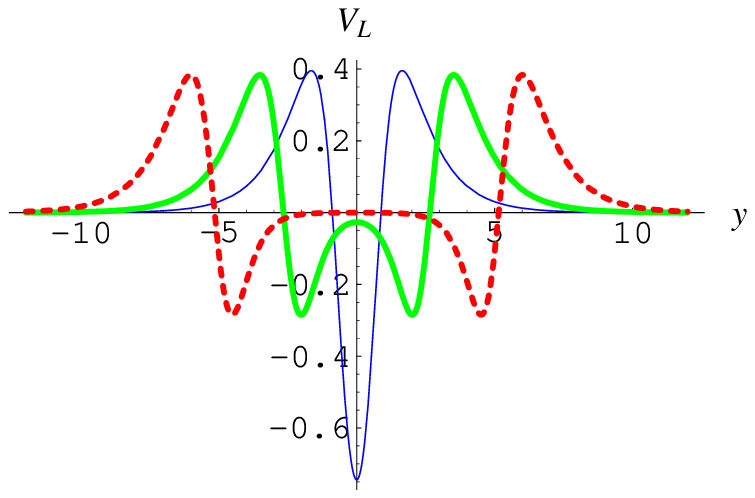}
\includegraphics[width=7cm]{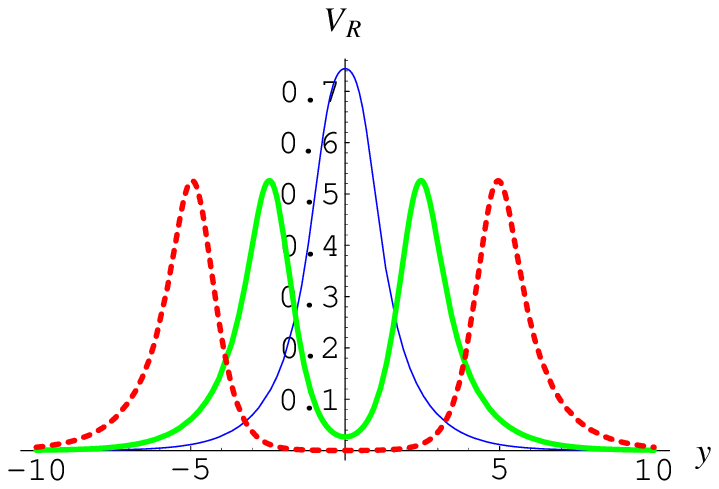}
\end{center}
\vskip -4mm
 \caption{The shape of the potentials $V_{L,R}$
  for the symmetric branes with $a=b$
  for different values of $\delta_0$ in $y$ coordinates.
 The parameters are set to $b=1,~v=1,~\beta=0,~\eta=1$,
 $\delta_0=10~(u=0.00018)$ for the red dashed lines,
 $\delta_0=5~(u=0.027)$ for the green thick lines, and
 $\delta_0=1~(u=1.47)$ for the blue thin lines. }
 \label{fig_VLVR_delta_0}
\end{figure}

\begin{figure}[htb]
\begin{center}
\includegraphics[width=7cm]{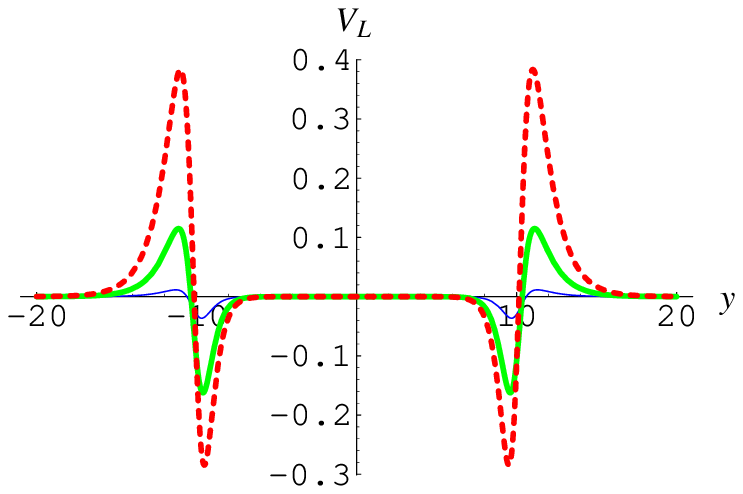}
\includegraphics[width=7cm]{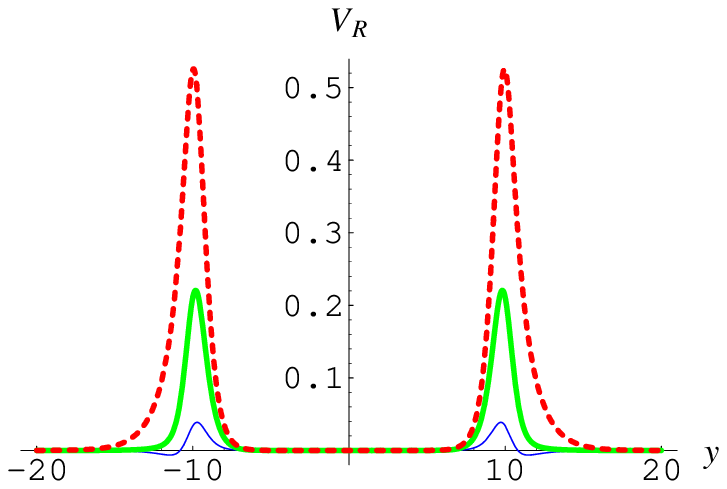}
\end{center}
\vskip -4mm
 \caption{The shape of the potentials $V_{L,R}$
  for the symmetric branes with $a=b$
  for different Yukawa coupling constant $\eta$ in $y$ coordinates.
 The parameters are set to $b=1,~v=1,~\beta=0,~\delta_0=20~(u=8.2\times10^{-9})$,
 $\eta_0=1$ for the red dashed lines,
 $\eta_0=0.5$ for the green thick lines, and
 $\eta_0=0.1$ for the blue thin lines. }
 \label{fig_VLVR_eta}
\end{figure}

\begin{figure}[htb]
\begin{center}
\includegraphics[width=7cm]{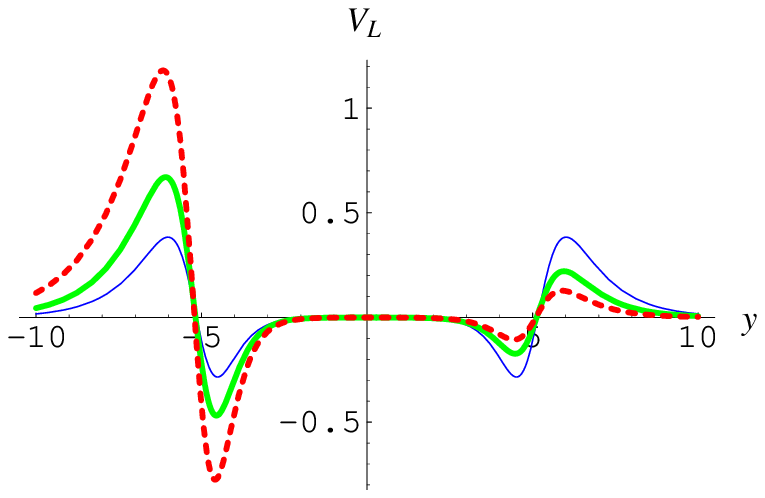}
\includegraphics[width=7cm]{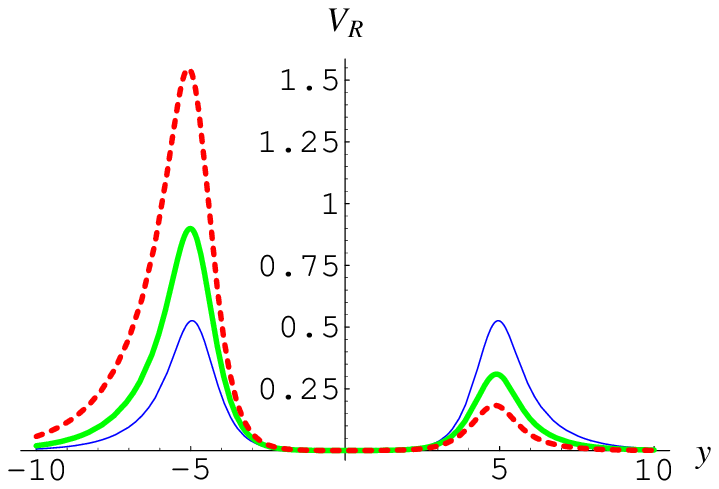}
\end{center}
\vskip -4mm
 \caption{The shape of the potentials $V_{L,R}$
  for the asymmetric branes with $a=b$
  for different values of the asymmetric factor $\beta$ in $y$ coordinates.
 The parameters are set to $b=1,~v=1,~\eta=1,~\delta_0=10~(u=0.00018)$,
 $\beta_0=0.1$ for the red dashed lines,
 $\beta_0=0.05$ for the green thick lines, and
 $\beta_0=0$ for the blue thin lines. }
 \label{fig_VLVR_beta}
\end{figure}

\subsection{The zero mode}
\label{secFermionZeroMode}

For positive $\eta$ and $b$,
we will show that the potential for left-handed fermions
could trap the left-handed fermion zero mode, which can be solved from
(\ref{CoupleEq1a}) by setting $m_0=0$:
\begin{equation}
 f_{L0}(z)
 \propto \exp\left(-\eta\int^z_0 dz'
     \text{e}^{A(z')}\phi(z')\right).
  \label{zeroModefL0CaseI}
\end{equation}
In order to check whether the zero mode can be localized on the brane, we
need to consider the normalizable problem of the solution. The
normalization condition for the zero mode (\ref{zeroModefL0CaseI}) is
\begin{eqnarray}
\int_{-\infty}^{\infty} dz \exp \left(-2\eta\int^z_0 dz'
     e^{A(z')}\phi(z')\right)<\infty.\label{eq:Norconditon-z}
\end{eqnarray}
Since we do not know the analytic expressions of the functions
$A(z)$ and $\phi(z)$ in $z$ coordinates, we have to
deal with the problem in $y$ coordinates. With the relation
(\ref{coordinateTransformation}), we can switch the
normalization condition (\ref{eq:Norconditon-z}) to the following one
in $y$ coordinates:
\begin{eqnarray}
\int_{-\infty}^{\infty} dy \exp\left( -A(y)-2\eta\int^y_0 dy'
\phi(y')\right)<\infty.\label{eq:Norconditon-y}
\end{eqnarray}
Now, it is clear that, because $\exp(-A(y))$ is divergent at
$y\rightarrow\pm\infty$, the introduction of the scalar-fermion
coupling is necessary in order to localize the fermion zero mode on the brane.
Since the functions $A(y)$ and $\phi(y)$ are smooth,
the normalization of the zero mode is decided by the asymptotic
characteristic of $\eta\phi(y)$ at $y\rightarrow\pm\infty$.

The integrand in (\ref{eq:Norconditon-y}) can be expressed explicitly as
\begin{eqnarray}
I_{1}
  &\propto& \exp\bigg[
    - \frac{2uv^2(u-c_0 \cosh (2bvy))}{9(u \cosh (2bvy)-c_0)^2}
    + \beta y \nonumber \\
  &&~~~~~ +\left(\frac{2v^2}{9}-\frac{\eta}{b}\right)\ln \left( u\cosh(2bvy)-c_{0} \right)
      \bigg ].
\end{eqnarray}
The asymptotic
characteristic of $I_1$ at $y\rightarrow\pm\infty$ is
\begin{eqnarray}
I_{1} \rightarrow\left\{\begin{array}{ll}
 \exp\left( -2v\big(\eta-\frac{2bv^2}{9}-\frac{\beta}{2v}\big) y\right)
        & ~$when$ ~y\rightarrow+\infty \\
 \exp\left( +2v\big(\eta-\frac{2bv^2}{9}+\frac{\beta}{2v}\big) y\right)
        & ~$when$ ~y\rightarrow-\infty
       \end{array} \right. .
\end{eqnarray}
So, the normalization condition of the zero mode turns out to be
\begin{eqnarray}
\eta>\frac{2bv^2}{9}+\left|\frac{\beta}{2v}\right|, \label{conditionCaseI}
\end{eqnarray}
which is simplified for the symmetric brane as $\eta>{2bv^2}/{9}$.
Provided the condition (\ref{conditionCaseI}), the zero mode of left-handed
fermions can be localized on the brane.
It can be seen that in order for the potential $V_L$ to localize the zero
mode of left-handed fermions for larger $v$, $b$ or the asymmetric
factor $\beta$, the stronger scalar-fermion coupling constant $\eta$ is required.
That is to say, the massless mode of left-handed fermion is
easier to be localized on the symmetric brane,
and the asymmetric factor $\beta$ makes it more difficult
to localize the zero mode.
A similar result was also found in Ref. \cite{Liu0907.0910}. However, this
is different from the situation of the zero modes of scalars and
vectors on symmetric and asymmetric de Sitter branes
\cite{LiuJCAP2009}, where increasing the asymmetric factor does
not change the number of the bound vector KK modes but would
increase that of the bound scalar KK modes, and the zero modes of
scalars and vectors are always localized on the de Sitter branes.
In Refs. \cite{KoleyCQG2005,LiuPRD2008},
it was shown that the zero mode of left-handed fermions
can also be localized on the brane in the background of
Sine-Gordon kinks provided similar condition as
(\ref{conditionCaseI}). The fermion zero mode cannot be
localized on the de Sitter brane with the same coupling
$F(\phi)=\phi$ \cite{LiuJCAP2009}.

The zero mode (\ref{zeroModefL0CaseI}) can be explicitly
written as a function of $y$:
\begin{equation}
 f_{L0}(z(y))
 \propto  \big[u\cosh(2bvy)-c_0 \big]^{-\frac{\eta}{2b}}.
  \label{zeroModefL0y}
\end{equation}
We note here that, although the zero mode $f_{L0}$ is an even function
in the $y$ coordinate, it is asymmetric in the $z$ coordinate
for the asymmetric brane because of the asymmetric warp factor.
In Fig. \ref{fig_ZeroMode_VL_DB1}, we plot the left-handed fermion
potential $V_L$ and the corresponding zero mode for the
symmetric and asymmetric branes in both the $y$ and $z$ coordinates.
We see that the zero mode is bound on the branes. It represents the
ground state of the Schr\"{o}dinger-like
equation (\ref{SchEqLeftFermion}) since it has no zero. Since
the ground state has the lowest mass square $m_0^2=0$, there is no
tachyonic left-handed fermion mode.
In fact, the differential equations (\ref{SchEqLeftFermion}) and
(\ref{SchEqRightFermion}) can be factorized as
\begin{eqnarray}
 \left[-\partial_z+\eta\;\text{e}^A F(\phi)\right]
 \left[\partial_z+\eta\;\text{e}^A F(\phi) \right]
 f_{Ln}(z) &=& m_n^2 f_{Ln}(z), \label{SchEqLeftFermion2}  \\
 \left[-\partial_z-\eta\;\text{e}^A F(\phi)\right]
 \left[\partial_z-\eta\;\text{e}^A F(\phi) \right]
 f_{Rn}(z) &=& m_n^2 f_{Rn}(z). \label{SchEqRightFermion2}
\end{eqnarray}
It can be shown that $m_n^2$ is zero or positive since the resulting
Hamiltonian can be factorized as the product of two operators which
are adjoints of each other. Hence the system is stable.

The zero mode $f_{L0}$ on both the
symmetric and asymmetric double branes is essentially constant
between the two sub-branes. This is different from the case of
gravity, where the gravitational zero mode $h_0$ on the
asymmetric double brane is strongly localized on the sub-brane
centered around the lower minimum of the potential.
In Ref. \cite{Liu0907.0910}, fermions on one-field-generating
symmetric and asymmetric
double branes were also studied, where the asymmetric
brane solution was constructed from a symmetric one with
another method presented in \cite{asymdSBrane2}.
It was found that the corresponding fermion zero mode $f_{L0}$
on both the double branes for the same Yukawa coupling $F(\phi)=\phi$
is also constant between the two sub-branes.

\begin{figure}[htb]
\begin{center}
\subfigure[$\beta=0$]{\label{fig_VL_ZeroMode_VG_DB1_a}
\includegraphics[width=7cm,height=4.5cm]{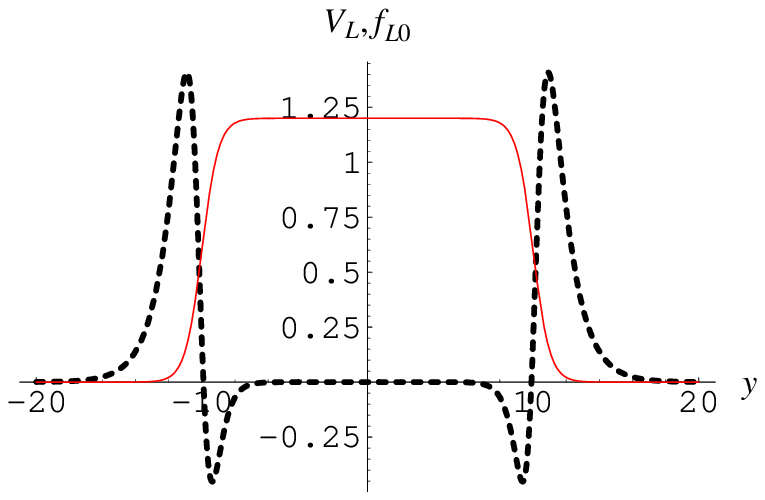}}
\subfigure[$\beta=0$]{\label{fig_VL_ZeroMode_VG_DB1_a}
\includegraphics[width=7cm,height=4.5cm]{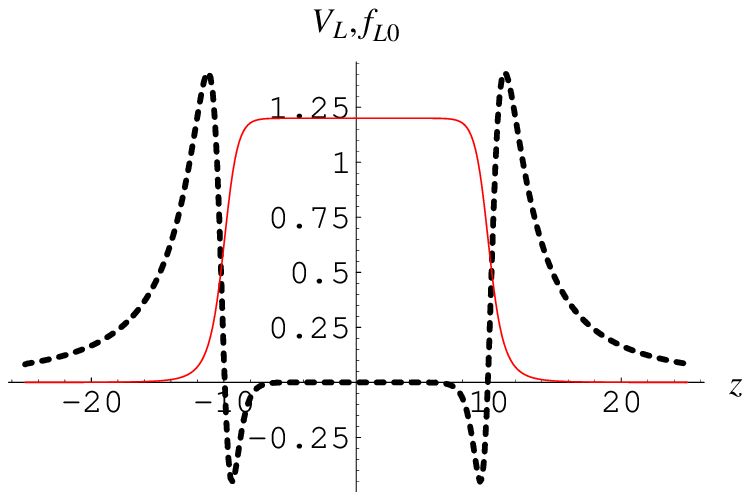}}
\subfigure[$\beta=1/16$]{\label{fig_VL_ZeroMode_y_as}
\includegraphics[width=7cm,height=4.5cm]{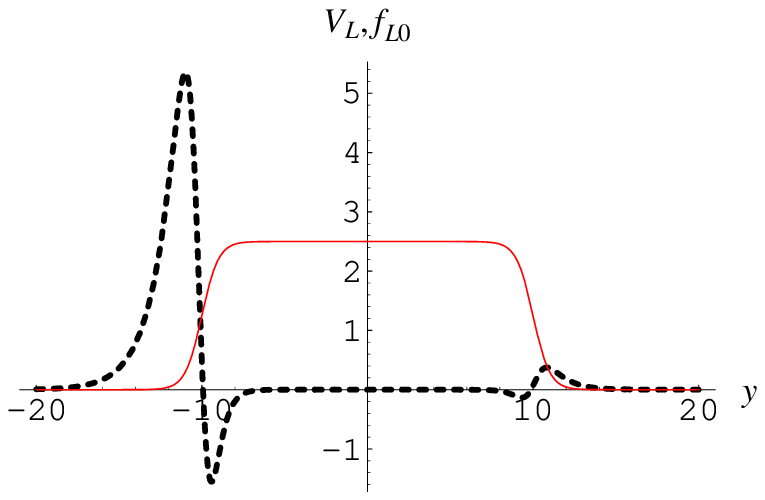}}
\subfigure[$\beta=1/16$]{\label{fig_VL_ZeroMode_z_as}
\includegraphics[width=7cm,height=4.5cm]{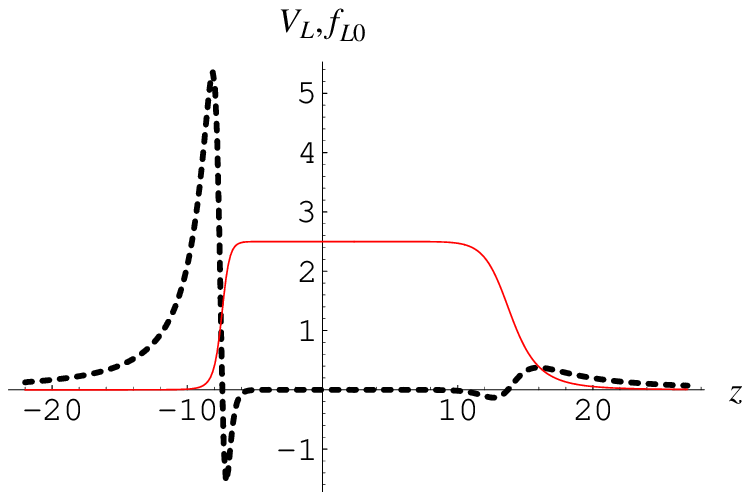}}
\end{center}
\vskip -4mm
 \caption{The potential $V_L$ (thick dashed lines)
 and the fermion zero mode $f_{L0}$ (red thin lines) for the symmetric
 ($\beta=0$, up) and asymmetric ($\beta=1/16$, down)
 double branes with $a=b$ in $y$ (left) and
 $z$ (right) coordinates.
 The parameters are set to $b=1,~v=1,~\delta_0=20,~\eta=2$.}
 \label{fig_ZeroMode_VL_DB1}
\end{figure}

\subsection{The massive KK modes and the resonances}
\label{secFermionMassiveMode}

The massive
modes will propagate along the extra dimension and those with lower
energy would experience an attenuation due to the presence of the
potential barriers near the location of the brane.

The potential $V_R$ is always positive near the brane location and
vanishes when far away from the brane. This shows that it could not
trap any bound fermions with right chirality and there is no zero
mode of right-handed fermions. However, the shape of the potential
is strongly dependent on the scalar-fermion coupling constant $\eta$. When $\eta$ increases to a certain value, the
potential will appear as a quasiwell around the brane location and the quasiwell would become deeper and deeper,
and as a result, this quasiwell will ``trap" some fermions; in fact, they are called
the quasilocalized or resonant fermions. Moreover, the number of resonances will
increase with the scalar-fermion coupling constant $\eta$. In
Ref. \cite{0901.3543}, a similar potential and resonances for left-
and right-handed fermions were found in background of two-field
generated thick branes with internal structure, which demonstrates
that a Dirac fermion could be composed from the left- and right-handed fermion resonance KK modes \cite{Liu0904.1785}.

We can investigate
the massive modes of fermions by solving numerically Eqs.
(\ref{SchEqLeftFermion}) and (\ref{SchEqRightFermion}).
For the set of parameters $b=v=1,~\beta=0,~\eta=2$ and $~\delta_0=20$, we find eight resonances
located at $m^2=\{0.02704$, $0.1072$, $0.2379$, $0.41533$, $0.6346$, $0.89013$, $1.1766$, $1.4955\}$ for both left- and right-handed
fermions (see Figs. \ref{fig_Fermion_VR_probability_P}(a) and \ref{fig_Fermion_VR_probability_P}(b)).

For $b=v=1,~\beta=0,~\eta=2$ and $~\delta_0=40$,
i.e., a brane with double width compared to the one with the same values of
$b,~v,~\beta,~\eta$ but $\delta_0=20$ (see Figs. \ref{fig_Fermion_VR_probability_P}(c) and \ref{fig_Fermion_VR_probability_P}(d)),
we find sixteen resonances for right-handed fermions,
and the mass spectrum of the resonances is calculated as
\begin{eqnarray}
  m^2 &=&\{0.0065, 0.0258, 0.0580, 0.1028, 0.1602,0.2299,  \nonumber\\
      && ~ 0.3116, 0.4051, 0.5100, 0.6260,0.7524,  \nonumber\\
      && ~ 0.8889, 1.0350, 1.1905, 1.3559, 1.5320 \}.
\end{eqnarray}

\begin{figure}[htb]
\begin{center}
\subfigure[]{\label{fig_Fermion_VR_M2_delta20}
\includegraphics[width=7cm,height=4.5cm]{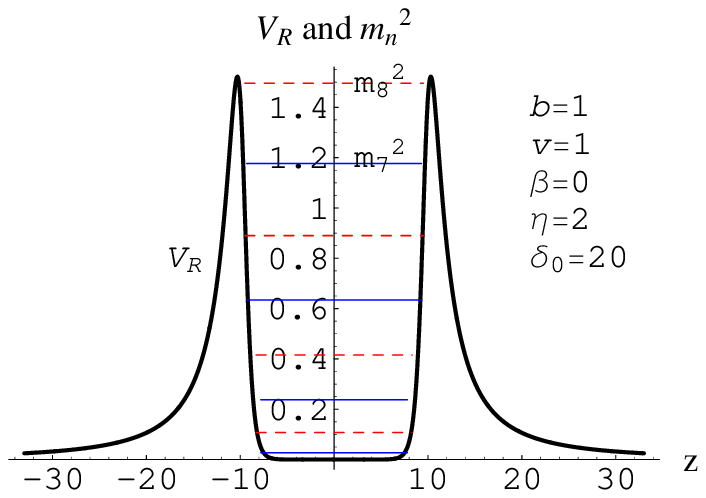}}
\subfigure[]{\label{fig_Fermion_PR_M2_delta20}
\includegraphics[width=7cm,height=4.5cm]{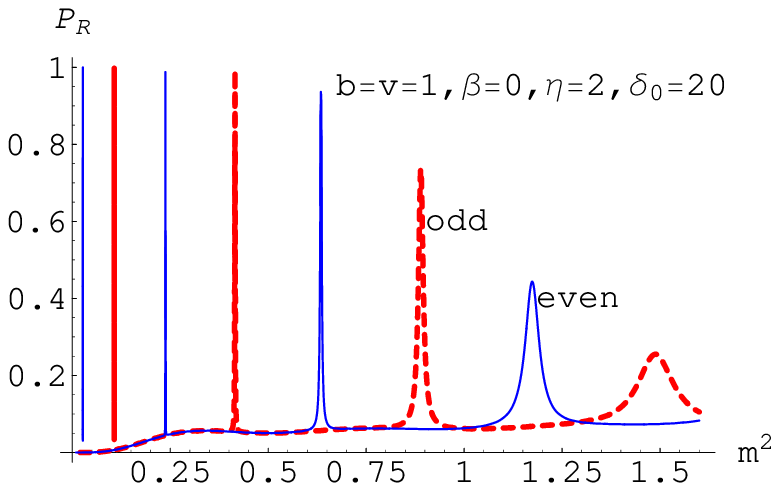}}
\subfigure[]{\label{fig_Fermion_VR_M2_delta40}
\includegraphics[width=7cm,height=4.5cm]{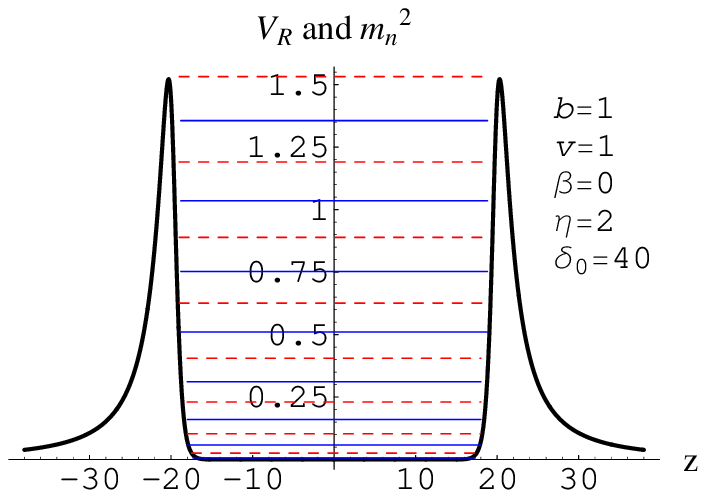}}
\subfigure[]{\label{fig_Fermion_PR_M2_delta40}
\includegraphics[width=7cm,height=4.5cm]{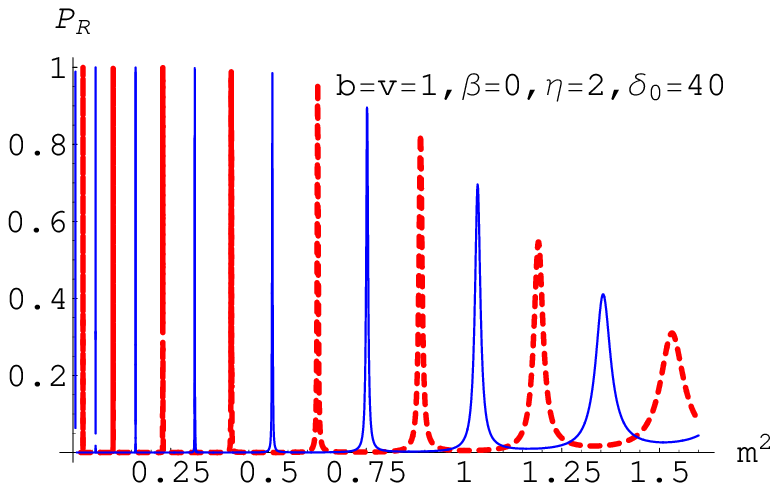}}
\end{center}
\vskip -4mm
  \caption{The potential $V_R(z)$, the resonance spectrum $m_n^2$, and
 the probability $P_R(m^2)$ for the right-handed fermion KK modes $f_{Rn}$.
 In the figures of $V_R(z)$ and $m_n^2$,
 $V_R(z)$ is denoted by black thick lines;
 $m_n^2$ for the odd and even resonance KK modes $f_{Rn}$ are denoted by red dashed and blue thin lines, respectively.
 In the figures of $P_R(m^2)$, the curves of $P_R(m^2)$ for odd and even
 modes $f_{Rn}$ are also denoted by red dashed lines and blue thin lines, respectively.
 }
 \label{fig_Fermion_VR_probability_P}
\end{figure}

We next analyze the lifetime of a fermion resonance.
First, we define the width $\Gamma=\Delta m$ of each resonant state as the
width at the half maximum of a resonant peak. In this
case, a massive fermion will disappear into the fifth dimension
after staying on the brane for some time $\tau \simeq \Gamma^{-1}$.
Thus, $\tau$ is called the lifetime of a fermion resonance
mentioned above. After numerical calculations, we
get a lifetime from each peak of the fermion resonance, which is
shown in detail in Fig. \ref{fig_Fermion_VR_lifetime}(a).
We find that the first peak is the most narrow one; accordingly, the lifetime of
this resonant state is the maximum, but it decays in exponential mode with the number $n$ increase,
so we get the conclusion that the KK
modes with a lower resonant state have a longer lifetime on the brane.
However, for a certain resonant state and brane width, the lifetime increases with the larger coupling constant $\eta$.
Also, from Fig. \ref{fig_Fermion_VR_lifetime}(b),
we can see that the total resonance number $N$ increases with
the coupling constant $\eta$ and it increases linearly with
the width of the double brane $\delta$.

\begin{figure}[htb]
\begin{center}
\subfigure[]{\label{fig_Fermion_VR_lifetime}
\includegraphics[width=7cm,height=4.5cm]{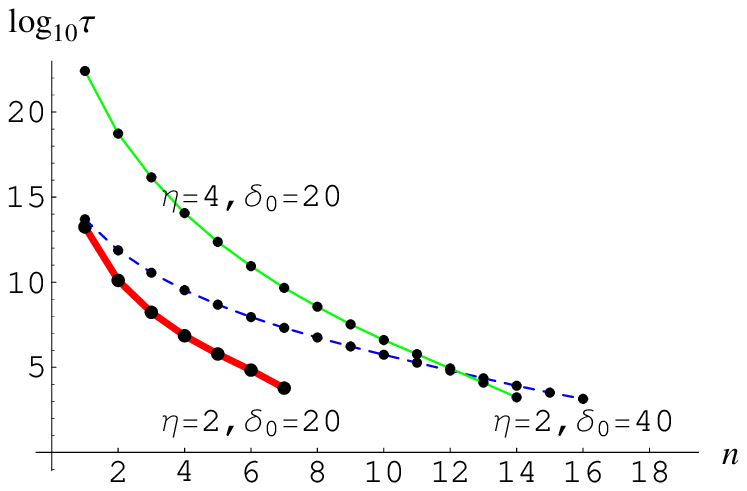}}
\subfigure[]{\label{fig_Fermion_VR_N_delta}
\includegraphics[width=7cm,height=4.5cm]{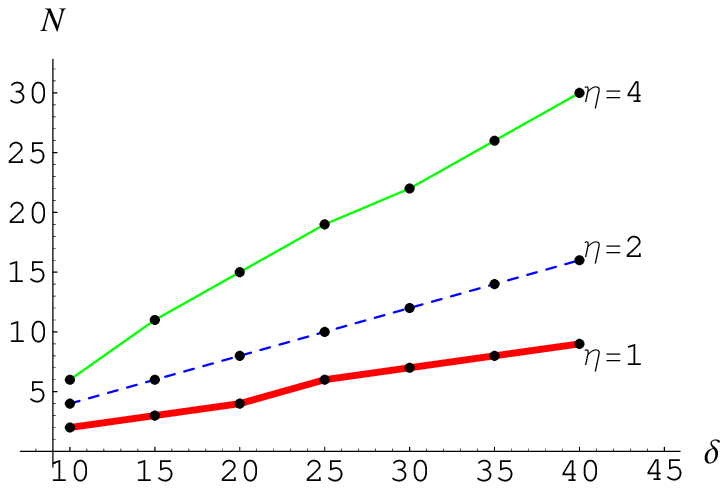}}
\end{center}
\vskip -4mm
 \caption{The relationship between $\text{log}_{10} \tau$ and $n$ (left), the total resonance number $N$  and the width of the double brane $\delta$
 (right). Here $\tau$ denotes the lifetime of a fermion resonance, and $n$ labels the order of the resonances.
 The parameters are set to $b=1,~v=1,~\beta=0$.}
 \label{fig_Fermion_VR_lifetime}
\end{figure}

We also use the transfer matrix method to calculate the resonance structure of fermion KK modes. For the symmetric case, the result is consistent with that of the relative possibility method, see Figs.~\ref{fig_Fermion_VR_probability_P} and \ref{fig_resonance_transmissivity_sym_Fermion}. For the asymmetric case, the result is shown in Fig.~\ref{fig_resonance_transmissivity_asym_Fermion}. Note that the resonance structures for $\beta$ and $-\beta$ are the same, because the potentials for $\beta$ and $-\beta$ have the same shape. From Figs.~\ref{fig_Fermion_VR_probability_P} and \ref{fig_resonance_transmissivity_asym_Fermion}, it can be seen that the
the number and the transmission coefficients of the fermion resonant KK modes decrease with the asymmetric factor $\beta$ rapidly. For $\beta=0.01$, the transmission coefficients of the fermion resonant KK modes with mass square smaller than 1.25 are almost around 0.3. For a larger $\beta=0.02$, the transmission coefficients of those resonances within the quasipotential well decrease to 0.05, which is much less than 1, the value of the symmetric brane case. So the fermion resonance structure is more sensitive to the asymmetric factor than that of the gravity one.

\begin{figure}[htb]
\begin{center}
\includegraphics[width=7cm,height=4.5cm]{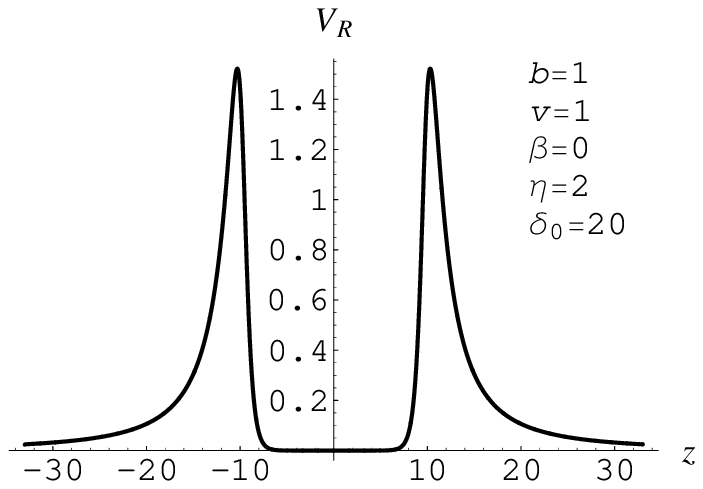}
\includegraphics[width=7cm,height=4.5cm]{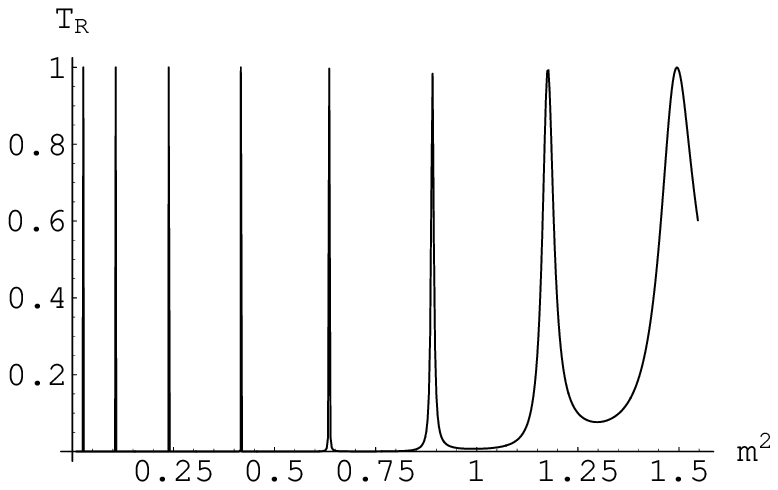}
\includegraphics[width=7cm,height=4.5cm]{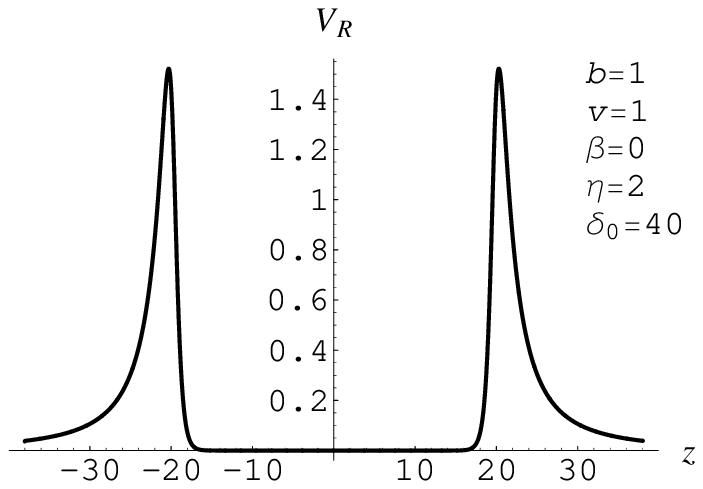}
\includegraphics[width=7cm,height=4.5cm]{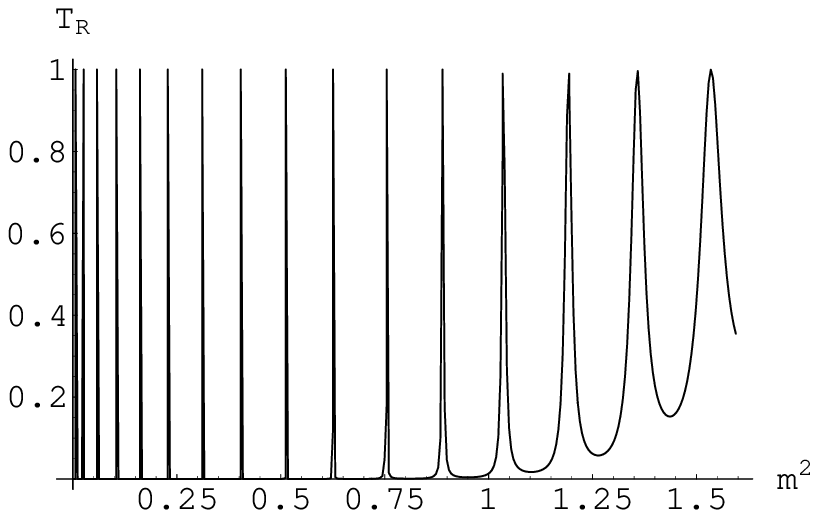}
\end{center}\vskip -4mm
 \caption{The potential $V_R(z)$ and the transmission coefficient $T_R$ of right-handed fermion KK modes on the symmetric Bloch brane.}
 \label{fig_resonance_transmissivity_sym_Fermion}
\end{figure}

\begin{figure}[htb]
\begin{center}
\includegraphics[width=7cm,height=4.5cm]{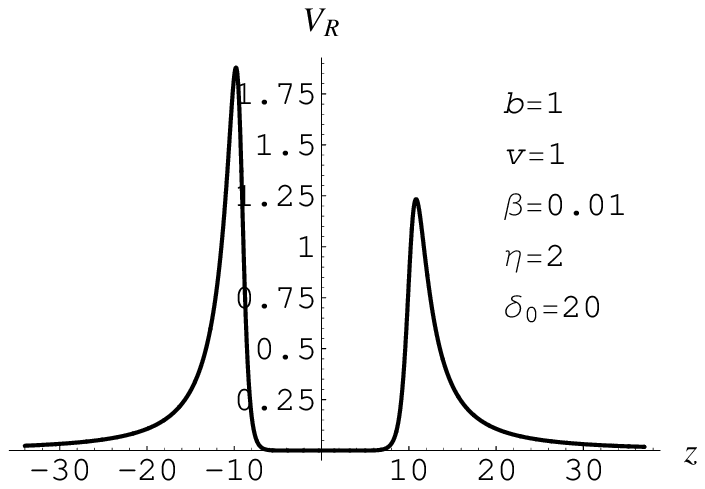}
\includegraphics[width=7cm,height=4.5cm]{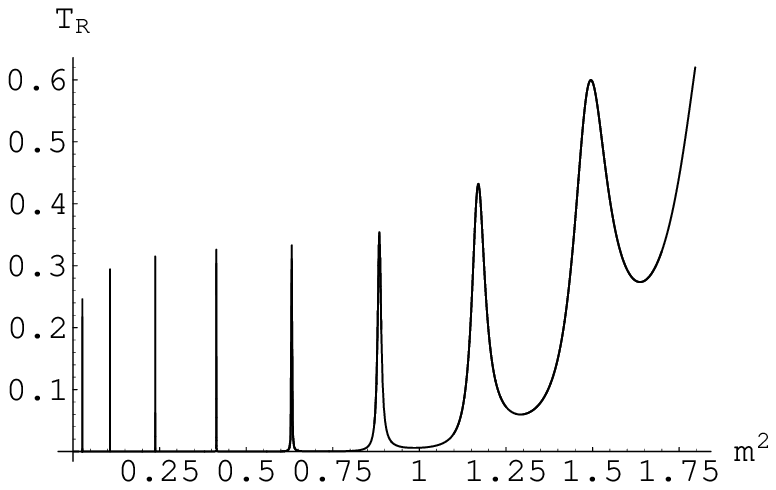}
\includegraphics[width=7cm,height=4.5cm]{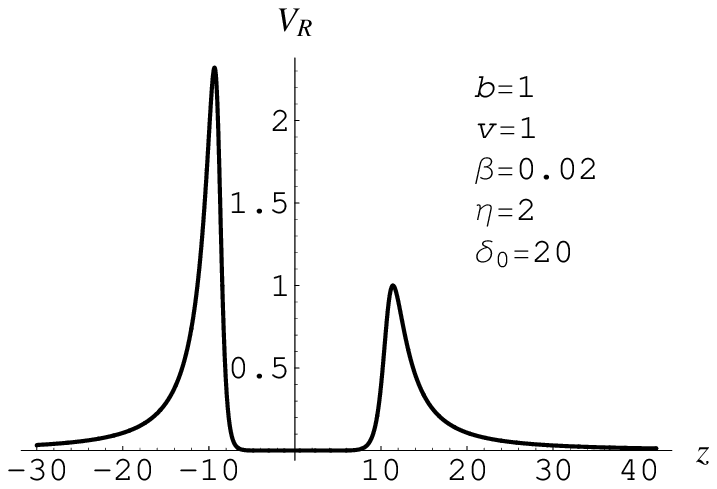}
\includegraphics[width=7cm,height=4.5cm]{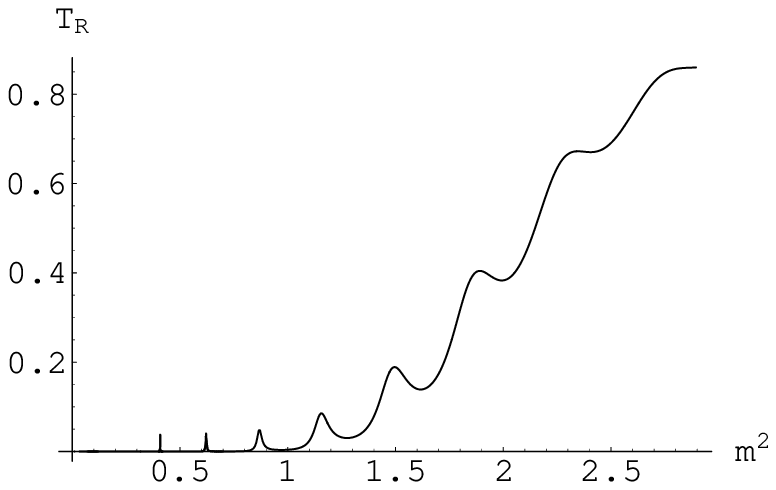}
\end{center}\vskip -4mm
 \caption{The potential $V_R(z)$ and the transmission coefficient $T_R$ of right-handed fermion KK modes on the asymmetric Bloch brane. }
 \label{fig_resonance_transmissivity_asym_Fermion}
\end{figure}

\section{Localization of the zero modes of the gravity and
fermion on the asymmetric Bloch double brane}
 \label{secStrongWeakBranes}

In Secs. \ref{SecGravityZeroMode} and \ref{secFermionZeroMode}, we have investigated the zero modes of the gravity and
fermion. However, the four-dimensional massless graviton and fermion
are not denoted by $h_0(z)$ and $f_0(z)$, respectively.
In fact, from the gravitational perturbation (\ref{pertubation_metric})
of the metric, it can be seen that the real tensor perturbation of the metric
is $\hat{h}_{\mu \nu}(x,z)$ but not $h_{\mu \nu}(x,z)$.
Therefore, from Eqs. (\ref{pertubation_metric}) and (\ref{decompsoseh_mu_nu}), it can be seen that
the four-dimensional massless graviton is presented by
\begin{equation}
 \hat{h}^{(0)}_{\mu \nu}(x,z)=e^{2A(z)}{h}^{(0)}_{\mu \nu}(x,z)
 = e^{ikx}\varepsilon_{\mu\nu}~e^{\frac{1}{2}A(z)}h_0(z),
\end{equation}
and the localization of the four-dimensional massless graviton
is characterized by $\hat{h}_{0}(z)=e^{\frac{1}{2}A(z)}{h}_{0}(z)$.
Similarly, the four-dimensional massless fermions with left chirality
and right chirality are presented by
$\psi_{L0}(x) \hat{f}_{L0}(z)=e^{-2A(z)}\psi_{L0}(x){f}_{L0}(z)$ and
$\psi_{R0}(x) \hat{f}_{R0}(z)=e^{-2A(z)}\psi_{R0}(x){f}_{R0}(z)$, respectively.
Hence the localization of the four-dimensional massless fermions
are characterized by $\hat{f}_{L0}(z)=e^{-2A(z)}{f}_{L0}(z)$
or $\hat{f}_{R0}(z)=e^{-2A(z)}{f}_{R0}(z)$.

The energy density $\rho$, the gravitational zero mode $\hat{h}_0$
and the fermion zero mode $\hat{f}_{L0}$ for the symmetric and asymmetric
double branes are plotted in Fig.~\ref{fig_EnergyDensity_Zeromodes}.
By comparing Figs.~\ref{fig_ZeroMode_VG_DB1}, \ref{fig_ZeroMode_VL_DB1}
and \ref{fig_EnergyDensity_Zeromodes}, we can find that
the localization properties of four-dimensional massless graviton and fermion in the symmetric
brane case read from two kinds of descriptions ($h_0$, $f_{L0}$ and $\hat{h}_0$, $\hat{f}_{L0}$) are the same;
i.e., the graviton and the fermion are localized between the two sub-branes.
However, for the asymmetric brane case, the results for the fermion caused by
the two kinds of descriptions are not the same. As mentioned above,
since the localization of the four-dimensional massless graviton
and fermion should be described by $\hat{h}_{0}(z)$ and $\hat{f}_{L,R0}(z)$,
respectively, we can conclude that the four-dimensional massless graviton
is localized on the left sub-brane, while the four-dimensional massless fermion
is localized on the right one.

\begin{figure}[htb]
\begin{center}
\includegraphics[width=7cm]{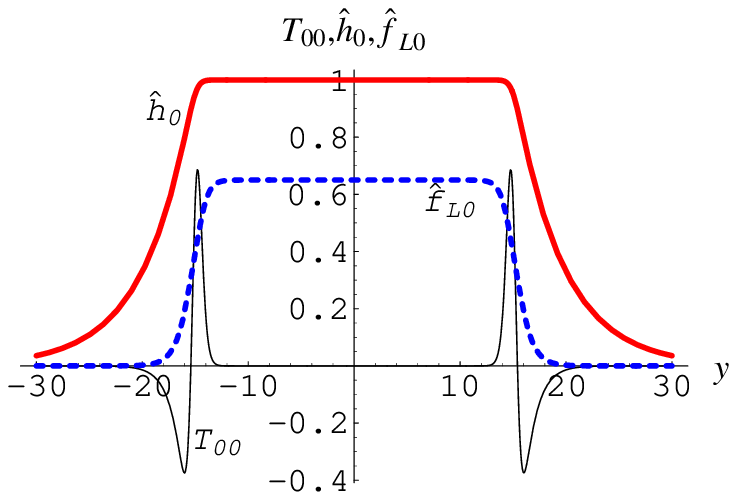}
\includegraphics[width=7cm]{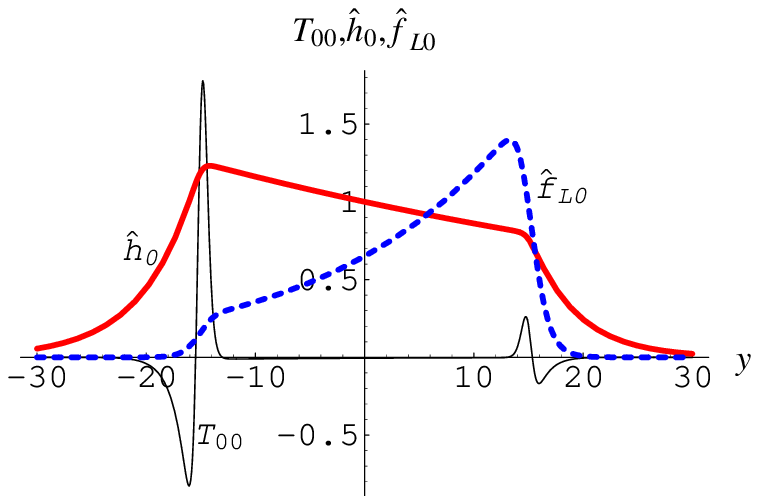}
\end{center}
\vskip -4mm
 \caption{The energy density $T_{00}$ (the thin lines),
  the gravitational zero mode $\hat{h}_0$ (the thick red lines)
  and the fermion zero mode $\hat{f}_{L0}$ (the dashed blue lines)
  for the symmetric ($\beta=0$, left) and asymmetric
  ($\beta=1/9$, right) double branes in $y$ coordinates.
  The other parameters are set to $b=1,~v=1,~\eta=2,~\delta_0=20$.
  The zero modes $\hat{h}_0$ and $\hat{f}_{L0}$
  are plotted with arbitrary normalization.}
 \label{fig_EnergyDensity_Zeromodes}
\end{figure}

\section{Discussions and conclusions}\label{secConclusion}

In this paper, by presenting the mass-independent potentials of the
KK modes in the corresponding Schr\"{o}dinger-like equations, we investigated the localization and mass spectra of
gravity and  spin-half fermion on
symmetric and asymmetric thick branes. We found that the zero mode of gravity can be localized on {the symmetric and asymmetric
branes}. For the case of fermion, there is no bound KK mode for both left- and
right-handed fermion zero modes without scalar-fermion
coupling ($\eta=0$). Hence, in order to localize the left- or right-handed
massless fermion KK mode on the brane, some kind
of Yukawa coupling should be introduced. In this paper, we considered the typical Yukawa coupling $\eta\bar{\Psi}\phi{\Psi}$, for which the localization conditions for the left-handed fermion zero modes are $\eta>\frac{2bv^2}{9}$ and $\eta>\frac{2bv^2}{9}+\left|\frac{\beta}{2v}\right|$ for the symmetric and asymmetric branes, respectively. So, it needs larger coupling constant in order to localize the fermion zero mode on the asymmetric brane. The zero mode of right-handed
fermion cannot be localized on the brane at the same condition.

The spectra of the KK modes are determined by the behavior of the potentials at
infinity. We found that all the potentials for the gravity and fermion KK modes turn to vanish at the boundary of extra dimension. Therefore, these potentials provide no mass gap to
separate the zero modes from the excited KK modes. Consequently, the spectra of the
gravity and the left- or right-handed fermion are composed of
a bound zero mode and a series of gapless continuous massive KK
modes.

For the case of a double brane, the potentials for the gravity and left-handed fermion KK modes almost vanish between the two sub-branes, and have two potential wells around the two sub-branes, while the potential for the right-handed fermion KK modes has a quasi potential well between the two sub-branes. For all of these potentials, there are two potential bars outside the two sub-branes. Such potentials result in some massive quasilocalized or resonant KK modes of gravity and fermion, which can stay on the branes for a certain time and then escape into the extra dimensions.
The resonant state with lower mass has a longer lifetime on the brane. For the case of gravity, the number of the resonances increases with the width of the double brane and the VEV of the scalar $\phi$.
For the case of left- and right-handed fermions, the lifetime of the $n$th resonance and the total number of the resonances increase with the scalar-fermion coupling constant $\eta$ and the width of the double brane.
We found that the mass spectra of left- and
right-handed fermion resonances are almost the same, which shows
that a Dirac fermion with a finite lifetime on the brane could be composed of the left- and right-handed fermion
resonance KK modes \cite{Liu0904.1785}.

For the asymmetric case, the fermion zero mode
is localized on one of the sub-branes, while the gravity zero mode is localized on another sub-brane.
The number and the transmission coefficients of the resonant KK modes of gravity of fermion will decrease with the asymmetric factor $\beta$. Just as in the symmetric brane case, the resonance with smaller mass square would have a longer lifetime.
For a larger enough asymmetric factor, while there is still a resonance structure in the $T-m^2$ picture, it does not denote resonance anymore.

At last, we give a comment on an important but more complex issue about the scalar spin-0
modes of the model, which would determine if the brane setup is stable or not.
This issue has been investigated in Refs. \cite{Shaposhnikov2005,George2011,Gherghetta2011} for one and $N$ real background scalar fields. It was found that the spin-0 degrees of freedom in the metric mix with the scalar perturbations. The zero modes of these perturbations were solved formally from the coupled Schr\"{o}dinger equations, and they can be used to construct a solution matrix. With the eigenvalues of the solution matrix, one can judge if a normalizable zero mode exists or not and how many normalizable negative mass modes exist in the spectrum. The case of two scalars were discussed in detail in Ref.~\cite{George2011}.
According to Ref.~\cite{George2011}, since the superpotential is used to generate the solution
in our model, there may exist
a scalar zero mode associated with the choice of integration
constant(s) of the solutions in Eqs. (\ref{DBbrane}). We will investigate this issue in future work.

Note-added. -Recently, we found another paper \cite{Cruz2013a} that also considered the resonances of gravity on a Bloch brane.

\section{ACKNOWLEDGMENT}

We thank the referee for his/her crucial comments and suggestions, which helped us to improve the original manuscript.
This work was supported by the National Natural Science Foundation of China (Grants No. 11075065 and 11375075), and
the Fundamental Research Funds for the Central Universities (Grant No. lzujbky-2013-18).

\end{document}